\documentclass[reqno]{amsproc}
\usepackage[latin1]{inputenc}
\usepackage{latexsym}
\usepackage[colorlinks, citecolor=blue,linkcolor=black]{hyperref}
\usepackage{cite}
\usepackage{xcolor}
\usepackage{mathrsfs,amstext,amsmath,amssymb,amsfonts,bm}
\newtheorem{definition}{Definition}
\newtheorem{example}[definition]{Example}

\newtheorem{remark}[definition]{Remark}

\makeatletter
\@addtoreset{definition}{section}
\@addtoreset{equation}{section}
\makeatother

\allowdisplaybreaks[4]
\def\6{\, {\rm d}}

\renewcommand{\epsilon}{\varepsilon}

\title{ Generalized super-$W_{1+\infty}$-$n$-algebra and Landau Problem }
\author{Fridolin Melong$^{\dagger}$ \and  Raimar Wulkenhaar$^{\ddagger}$}
\address[$\dagger$]{
	International Chair in Mathematical Physics and Applications,
	ICMPA-UNESCO Chair, \newline University of Abomey-Calavi, 072 BP 50, Cotonou, Rep. of Benin, {\itshape e-mail:} \normalfont
\texttt{fridolin.melong@uni-muenster.de  with copy to (fridomelong@gmail.com)}}
\address[$\ddagger$]{Mathematisches Institut der
	Universit\"at M\"unster \newline
	Einsteinstr.\ 62, 48149 M\"unster, Germany, 
	{\itshape e-mail:} \normalfont
	\texttt{raimar@math.uni-muenster.de}}
\begin{document}	
	\begin{abstract}
		We investigate the $\mathcal{R}(p,q)$-super $n$-bracket and study their properties such that the  generalized super Jacobi identity (GJSI). Furthermore, from the $\mathcal{R}(p,q)$-operators in a Supersymmetric Landau problem, we furnish  the
		$\mathcal{R}(p,q)$-super $W_{1+\infty}$ $n$-algebra which obey the generalized super Jacobi identity (GSJI) for $n$ even.  Also,  we derive  the $\mathcal{R}(p,q)$-super $W_{1+\infty}$ sub-$2n$-algebra and deduce particular cases induced by quantum algebras existing in the literature.
		\\
		{\noindent
			{\bf Keywords.  Quantum algebra, conformal and W symmetry, $n$-algebra, Supersymmetric Landau problem.}}
	\end{abstract}	
\maketitle
\tableofcontents
\section{Introduction}
The notion of 3-bracket for the 
Hamiltonian dynamics and  the higher order algebraic structures
with  applications in physics was introduced by Nambu\cite{N}.
 Recently the infinite-dimensional
n-algebras have received considerable attention \cite{CFZ,CKJ,AF,CWZ,DJWYW,YDLZZ}. As an
important infinite-dimensional algebra, the $W_{\infty}$ algebra
has been studied intensely from a variety of viewpoints.
Based on the structure of $W_{\infty}$ algebra, Chakraborty et
{\it al}\cite{CKJ} investigated the $W_{\infty}$ 3-algebra. By applying a double
scaling limit on the generators, they obtained a $W_{\infty}$ 3-
algebra, which satisfies the so-called fundamental identity
(FI). The super $W_{\infty}$ 3-algebra was described and their realization in terms of the super Nambu-Poisson
bracket was given in \cite{CWZ}. Also,  the $W_{1+\infty}$ 3-algebra was
studied and the  relation between  the integrable systems was established
in the framework of Nambu mechanics \cite{CWWWZ}.

Furthermore, the $W_{1+\infty}$ algebra has many important
applications in the quantum Hall effect and specially in the free electron theory of Landau levels. Moreover, the Supersymmetric extension of the $W_{1+\infty}$ algebra
appears in the Supersymmetric Landau problem. For more details see \cite{CTZ1,K,CTZ,IKS,A,ZDYWZ,H1}.

Morover, the  generalization of the  Heisenberg algebras named $\mathcal{R}(p,q)$-quantum algebras was introduced in \cite{HB1}. These  generalize deformed oscillator algebras well known in the physics literature. 
It is well known that, the $W_{1+\infty}$ algebra is viewed as  generalization of the Virasoro algebra. Recently, the generalizations of the Virasoro algebra and their sub-algebra from the point of view of  quantum algebra of one and multi-parameter deformations were investigated (see \cite{CZ,WYLWZ,CJ,HM,HMM,melong2022,melongwulkenhaar,melong2024}).  

Besides, the generalization of the $W_{\infty}$ $n$-algebra, elliptic hermitian matrix model and a toy model from the generalized quantum algebra were investigated and studied\cite{melongRaimar}.

 The aim of this work is to construct the $\mathcal{R}(p,q)$-super $n$-algebra and to characterize
the super $W_{\infty}$ $n$-algebra related to the Supersymmetric Landau
problem by the quantum algebra \cite{HB1}.

This paper is organized as follows. In section $2$, we construct the $\mathcal{R}(p,q)$-$W_{\infty}$-$3$-algebra by the limit of the $\mathcal{R}(p,q)$-super-high-order Virasoro algebra. Section $3$ is reserved to build  the generalization of the super $n$-algebra generated by the quantum algebra \cite{HB1}. In section $4$, we characterize  the $\mathcal{R}(p,q)$-super-$W_{\infty}$ algebra in the framework of the Supersymmetric Landau problem. In section $5$, we deduce particular cases of our results induced by quantum algebras in the literature. Finally,  we end with
the concluding remarks in section $6$.

\section{$\mathcal{R}(p,q)$-$W_{\infty}$ $3$-algebra}
In this section, we build the $W_{\infty}$ $3$-algebra induced by the quantum algebra \cite{HB1}. For that, we construct the $\mathcal{R}(p,q)$-super high-order Virasoro algebra and deduce the $\mathcal{R}(p,q)$-super high-order 3-algebra. Particular cases are investigated. 

We consider $ p$ and $q,$ two positive real numbers such that $ 0<q<p< 1,$ and a 
meromorphic function $\mathcal{R}$ defined on $\mathbb{C}\times\mathbb{C}$ by \cite{HB}: \begin{eqnarray}\label{r10}
\mathcal{R}(s,t)= \sum_{u,v=-l}^{\infty}r_{uv}s^u\,t^v,
\end{eqnarray}
where $r_{uv}$ are complex numbers, $l\in\mathbb{N}\cup\left\lbrace 0\right\rbrace,$ $\mathcal{R}(p^n,q^n)>0,  \forall n\in\mathbb{N},$ and $\mathcal{R}(1,1)=0$ by definition.
The bidisk $\mathbb{D}_{R}$ is defined by:  \begin{eqnarray*}
	\mathbb{D}_{R}
	&=&\left\lbrace a=(a_1,a_2)\in\mathbb{C}^2: |a_j|<R_{j} \right\rbrace,
\end{eqnarray*}
where $R$ is the convergence radius of the series (\ref{r10}) defined by Hadamard formula \cite{TN}:
$$	\lim\sup_{s+t \longrightarrow \infty} \sqrt[s+t]{|r_{st}|R^s_1\,R^t_2}=1.
$$
We also consider $\mathcal{O}(\mathbb{D}_{R})$ the set of holomorphic functions defined on $\mathbb{D}_{R}.$ We have:
\begin{itemize}
	\item 
 the  $\mathcal{R}(p,q)$-deformed numbers  \cite{HB}:
\begin{eqnarray}\label{rpqnumber}
[n]_{\mathcal{R}(p,q)}:=\mathcal{R}(p^n,q^n),\quad n\in\mathbb{N}\cup\{0\},
\end{eqnarray}
\item the
$\mathcal{R}(p,q)$-deformed factorials
\begin{eqnarray*}\label{s0}
	[n]!_{\mathcal{R}(p,q)}:=\left \{
	\begin{array}{l}
		1\quad\mbox{for}\quad n=0\\
		\\
		\mathcal{R}(p,q)\cdots\mathcal{R}(p^n,q^n)\quad\mbox{for}\quad n\geq 1,
	\end{array}
	\right .
\end{eqnarray*}
\item the  $\mathcal{R}(p,q)$-binomial coefficients
\begin{eqnarray}
\mathbb{C}^{k}_{n} := \frac{[n]!_{\mathcal{R}(p,q)}}{[k]!_{\mathcal{R}(p,q)}[n-k]!_{\mathcal{R}(p,q)}},\quad  n\geq k.
\end{eqnarray}
\item the linear operators defined on the set   $\mathcal{O}(\mathbb{D}_{R})$\cite{HB1}:
\begin{align*}
	\;Q:\psi\longmapsto Q\psi(z):&= \psi(qz),\\
	\; P:\psi\longmapsto P\psi(z):&=\psi(pz),\\
	\;\partial_{p,q}:\psi\longmapsto \partial_{p,q}\psi(z):&=\frac{\psi(pz)-\psi(qz)}{z(p-q)},
\end{align*}
  \item the $\mathcal{R}(p,q)$-derivative 
\begin{eqnarray*}\label{r5}
\mathcal{D}_{\mathcal{R}( p,q)}:=\mathcal{D}_{p,q}\frac{p-q}{P-Q}\mathcal{R}( P,Q)=\frac{p-q}{p^{P}-q^{Q}}\mathcal{R}(p^{P},q^{Q})\mathcal{D}_{p,q}.
\end{eqnarray*}
\end{itemize}
Let us take the $\mathcal{R}(p,q)$-operators of the super high-order Virasoro algebra  given as follows:
\begin{align}
\,\mathcal{L}_m^s&=(-1)^s{\lambda}^{s-\frac{1}{2}}\, z^{m+s}\mathcal{D}_{\mathcal{R}(p,q)}^s\label{Rqgen1},\\
\,\bar{\mathcal{L}}_m^s&=(-1)^{s}{\lambda}^{s+\frac{3}{2}}\,z^{m+s}\theta\,\hat{\mathcal{D}}_{\mathcal{R}(p,q)}\,\mathcal{D}_{\mathcal{R}(p,q)}^s\label{Rqgen2},\\
\,\mathcal{H}_r^{\alpha+\frac{1}{2}}&=(-1)^{\alpha +1}\lambda^{\alpha+\frac{1}{2}}\, z^{r
	+\alpha}\,\hat{\mathcal{D}}_{\mathcal{R}(p,q)}\,\mathcal{D}_{\mathcal{R}(p,q)}^\alpha\label{Rqgen3},\\
\,\bar{\mathcal{H}}_r^{\alpha+\frac{1}{2}}&=(-1)^{\alpha+1}
\lambda^{\alpha+\frac{1}{2}}\, z^{r+\alpha}\theta\,\mathcal{D}_{\mathcal{R}(p,q)}^\alpha\label{Rqgen4},
\end{align}
where $ r, \alpha\in \mathbb{Z}_+$, $m, r\in \mathbb{Z}$, $\lambda$ is an arbitrary
parameter,  $\mathcal{D}_{\mathcal{R}(p,q)}$ and $\hat{\mathcal{D}}_{\mathcal{R}(p,q)}$
are the derivatives with respect to $z$ and $\theta$, respectively.

We consider the commutation relation of $U$ and $V$ defined by:
\begin{eqnarray}\label{Rqbracket}
\left[U,V\right]=UV-(-1)^{|U||V|}VU,
\end{eqnarray}
$|U|$ and $|V|$ are the parity of $U$ and $V$, respectively.

Then, the operators \eqref{Rqgen1},\eqref{Rqgen2},\eqref{Rqgen3}, and \eqref{Rqgen4} satisfy the following commutation relations:
\begin{align}\label{Rqshov1}
[\mathcal{L}_m^s, \mathcal{L}_n^r]&=\mathcal{K}(P,Q)p^{k}\,\sum_{k=0}^{s} (-1)^k\lambda^{k-\frac{1}{2}}\,q^{(s-k)(n+r)}\,\mathbb{C}_s^k\,
\mathbb{A}^{k}_{n+r}\, \mathcal{L}_{m+n}^{s+r-k}\nonumber\\
&-\mathcal{K}(P,Q)p^{k}\,\sum_{k=0}^r(-1)^k\lambda^{k-\frac{1}{2}}\,q^{(r-k)(m+s)}\,
\mathbb{C}_r^k\, \mathbb{A}^k_{m+s}\,\mathcal{L}_{m+n}^{s+r-k},
\end{align}
\begin{align}\label{Rqshov2}
\left[\mathcal{L}_m^s, \bar{ \mathcal{L}}_n^r\right]&=\mathcal{K}(P,Q)p^{k}\,\sum_{k=0}^{s}(-1)^k\lambda^{k-\frac{1}{2}}\,q^{(s-k)(n+r)}\, \mathbb{C}_s^k\,
\mathbb{A}^{k}_{n+r}\, \bar{\mathcal{L}}_{m+n}^{s+r-k}\nonumber\\
&-\mathcal{K}(P,Q)p^{k}\,\sum_{k=0}^r(-1)^k\lambda^{k-\frac{1}{2}}\,q^{(r-k)(m+s)}\,\mathbb{C}_r^k\, \mathbb{A}^{k}_{m+s}\,
\bar{\mathcal{L}}_{m+n}^{s+r-k},
\end{align}
\begin{align}\label{Rqshov3}
\left[\mathcal{L}_m^s, \mathcal{H}_r^{\alpha+\frac{1}{2}}\right]&=\mathcal{K}(P,Q)\sum_{k=0}^{s}(-1)^k
\,\lambda^{k-\frac{1}{2}}p^{k}\,q^{(s-k)(r+\alpha)}\,\mathbb{C}_s^k \mathbb{A}^{k}_{r+\alpha}
\mathcal{H}_{m+r}^{\alpha+s-k+\frac{1}{2}}\nonumber\\
&-\mathcal{K}(P,Q)\sum_{k=0}^\alpha\,(-1)^k\,
\lambda^{k-\frac{1}{2}}p^{k}\,q^{(\alpha-k)(m+s)}\,\mathbb{C}_\alpha^k\, \mathbb{A}^k_{m+s}\,
\mathcal{H}_{m+r}^{\alpha+s-k+\frac{1}{2}},
\end{align}
\begin{align}\label{Rqshov4}
\left[\mathcal{L}_m^s, \bar{\mathcal{H}}_r^{\alpha+\frac{1}{2}}\right]&=\mathcal{K}(P,Q)\sum_{k=0}^{s}(-1)^k\,
\lambda^{k-\frac{1}{2}}\,p^{k}\,q^{(s-k)(r+\alpha)}\, \mathbb{C}_s^k\, \mathbb{A}^{k}_{r+\alpha}\, \bar
{\mathcal{H}}_{m+r}^{\alpha+s-k+\frac{1}{2}}\nonumber\\
&-\mathcal{K}(P,Q)\sum_{k=0}^\alpha\,(-1)^k
\lambda^{k-\frac{1}{2}}\,p^{k}\,q^{(\alpha-k)(m+s)}\,\mathbb{C}_\alpha^k \,\mathbb{A}^k_{m+s}\, \bar
{\mathcal{H}}_{m+r}^{\alpha+s-k+\frac{1}{2}},
\end{align}
\begin{align}\label{Rqshov5}
\left[\bar{\mathcal{L}}_m^s, \mathcal{H}_r^{\alpha+\frac{1}{2}}\right]=-\mathcal{K}(P,Q)\sum_{k=0}^{\alpha}
(-1)^{k}\lambda^{k+\frac{3}{2}}\,p^{k}\,q^{(\alpha-k)(m+s)}\, \mathbb{C}_\alpha^k
\mathbb{A}^k_{m+s}\,\mathcal{H}_{m+r}^{\alpha+s-k+\frac{1}{2}},
\end{align}
\begin{align}\label{Rqshov6}
\left[\bar {\mathcal{L}}_m^s, \bar {\mathcal{L}}_n^r\right]&=\mathcal{K}(P,Q)\sum_{k=0}^{s}\, (-1)^{k}\,\lambda^{k+\frac{3}{2}}\,p^{k}\,q^{(s-k)(n+r)}\,
\mathbb{C}_s^k\, \mathbb{A}^k_{n+r}\,  \bar {\mathcal{L}}_{m+n}^{s+r-k}\nonumber\\
&-\mathcal{K}(P,Q)\sum_{k=0}^r\,(-1)^{k}\,
\lambda^{k+\frac{3}{2}}\,p^{k}\,q^{(r-k)(m+s)}\,\mathbb{C}_r^k\, \mathbb{A}^k_{m+s}\, \bar {\mathcal{L}}_{m+n}^{s+r-k},
\end{align}
\begin{eqnarray}\label{Rqshov7}
[\bar {\mathcal{L}}_m^s, \bar {\mathcal{H}}_r^{\alpha+\frac{1}{2}}]=\mathcal{K}(P,Q)\sum_{k=0}^{s}\,(-1)^{k}\,\lambda^{k+\frac{3}{2}}\,p^{k}\,q^{(s-k)(r+\alpha)}\,
\mathbb{C}_s^k\, \mathbb{A}^{k}_{r+\alpha}\, \bar
{\mathcal{H}}_{m+r}^{\alpha+s-k+\frac{1}{2}},
\end{eqnarray}
\begin{align}\label{Rqshov8}
\left[\mathcal{H}_r^{\alpha+\frac{1}{2}}, \bar
{\mathcal{H}}_s^{\beta+\frac{1}{2}}\right]&=\mathcal{K}(P,Q)\sum_{k=0}^{\alpha}
(-1)^{k}\,\lambda^{k+\frac{3}{2}}\,p^{k}\,q^{(\alpha-k)(s+\beta)} \mathbb{C}_\alpha^k\, \mathbb{A}^k_{s+\beta}\,
\mathcal{L}_{r+s}^{\alpha+\beta-k}\nonumber\\&+\mathcal{K}(P,Q)\sum_{k=0}^{\beta}(-1)^k\,
\lambda^{k-\frac{1}{2}}\,p^{k}\,q^{(\beta-k)(r+\alpha)}\,
\mathbb{C}_\beta^k\,\mathbb{A}^k_{r+\alpha}\,\bar {\mathcal{L}}_{r+s}^{\alpha+\beta-k}\nonumber\\
&-\mathcal{K}(P,Q)\sum_{k=0}^{\alpha}\,(-1)^k\, \lambda^{k-\frac{1}{2}}\,p^{k}\,q^{(\alpha-k)(s+\beta)}\,
\mathbb{C}_\alpha^k\,\mathbb{A}^k_{s+\beta}\, \bar {\mathcal{L}}_{r+s}^{\alpha+\beta-k},
\end{align}
\begin{eqnarray}\label{Rqshov9}
\left[\bar{\mathcal{H}}_r^{\alpha+\frac{1}{2}}, \bar
{\mathcal{H}}_s^{\beta+\frac{1}{2}}\right]=0,
\end{eqnarray}
where 
\begin{eqnarray}
\mathbb{A}_n^k:=
\left\{\begin{array}{cc}
[n]_{\mathcal{R}(p,q)}[n-1]_{\mathcal{R}(p,q)}
\cdots[n-k+1]_{\mathcal{R}(p,q)},& k\leqslant n,\\
0,                  &k>n.\end{array}\right. 
\end{eqnarray}
\begin{remark} It is worth to note that:
	\begin{enumerate}
		\item[(i)] Setting $\lambda=1,$ we obtain another form of the $\mathcal{R}(p,q)$-super-high-order Virasoro algebra.
		\item[(ii)] 
		Taking $\mathcal{R}(x)=\frac{x-1}{q-1}$  in the relations
		\eqref{Rqshov1} to \eqref{Rqshov9}, we  obtain  the  commutation relations of the $q$-super high-order Virasoro algebra which are different from the results
		derived by Zha and Zhao \cite{zz}.
		\item[(iii)] Putting $\lambda=1$, we deduce the results given by \cite{CWZ}.
	\end{enumerate}
\end{remark}

Now, we define a $\mathcal{R}(p,q)$-super 3-bracket by:
\begin{eqnarray}\label{Rqbracket1}
\left[U, V, W\right]=\left[U, V\right]W+(-1)^{|U|(|V|+|W|)}\left[V, W\right]U+(-1)^{|W|(|U|+|V|)}\left[W, U\right]V,
\end{eqnarray}
where the commutator $[\ , \ ]$ is given  by the relation \eqref{Rqbracket}.

Introducing the $\mathcal{R}(p,q)$-operators \eqref{Rqgen1}-\eqref{Rqgen4} in the  generalized ternary commutator \eqref{Rqbracket1},
we can derive the $\mathcal{R}(p,q)$-super high-order Virasoro 3-algebra.
Due to the large numbers of  3-algebra relations, we restrict to five commutation relations in the sequel.
\begin{align}\label{Rqlim1}
\left[\mathcal{L}_m^s, \mathcal{L}_n^r, \mathcal{L}_k^h\right] &=\mathcal{K}(P,Q)\bigg[\bigg(\sum_{i=0}^{s}p^{i}\,q^{(s-i)(n+r)}\, \mathbb{C}_s^i\,
\mathbb{A}^i_{n+r}-\sum_{i=0}^rp^{i}\,q^{(r-i)(m+s)}\, \mathbb{C}_r^i\, \mathbb{A}^i_{m+s}\bigg)\nonumber\\
&\times\sum_{j=0}^{s+r-i}p^{j}\,q^{(s+r-i-j)(k+h)}\,
\mathbb{C}_{s+r-i}^j\, \mathbb{A}^j_{k+h}
+\bigg(\sum_{i=0}^{r}p^{i}\,q^{(r-i)(k+h)}\, \mathbb{C}_r^i\,
\mathbb{A}^i_{k+h}\nonumber\\
&-\sum_{i=0}^hp^{i}\,q^{(h-i)(n+r)}\,  \mathbb{C}_h^j\, \mathbb{A}^i_{n+r}\bigg)\sum_{j=0}^{r+h-i}p^{j}\,q^{(r+h-i-j)(m+s)}\,   \mathbb{C}_{r+h-i}^j\, \mathbb{A}^j_{m+s}\nonumber\\
&
+\bigg(\sum_{i=0}^{h}p^{i}\,q^{(h-i)(m+s)}\,\mathbb{C}_h^i\, \mathbb{A}^i_{m+s}-\sum_{i=0}^sp^{i}\,q^{(s-i)(k+h)}\, \mathbb{C}_s^i\, \mathbb{A}^i_{k+h}\bigg)\nonumber\\
&\times\sum_{j=0}^{h+s-i}p^{j}\,q^{(h+s-i-j)(n+r)}\,  \mathbb{C}_{h+s-i}^j \mathbb{A}^j_{n+r}\bigg]
(-1)^{i+j}\lambda^{i+j-1} \mathcal{L}_{m+n+k}^{s+r+h-i-j},
\end{align}
\begin{align}\label{Rqlim2}
\left[\mathcal{L}_m^s, \mathcal{L}_n^r, \bar{\mathcal{L}}_k^h\right]
&=\mathcal{K}(P,Q)\bigg[\bigg(\sum_{i=0}^{s}p^{i}\,q^{(s-i)(n+r)}\,\mathbb{C}_s^i \mathbb{A}^i_{n+r}-\sum_{i=0}^rp^{i}\,q^{(r-i)(m+s)} \mathbb{C}_r^i
\mathbb{A}^i_{m+s}\bigg)\nonumber\\&\times\sum_{j=0}^{s+r-i}p^{j}\,q^{(s+r-i-j)(k+h)}\, \mathbb{C}_{s+r-i}^j\,\mathbb{A}^j_{k+h}
+\bigg(\sum_{i=0}^{r}p^{i}\, \mathbb{C}_r^i\,q^{(r-i)(k+h)}\, \mathbb{A}^i_{k+h}\nonumber\\
&-\sum_{i=0}^hp^{i}\,q^{(h-i)(n+r)}\, \mathbb{C}_h^i\, \mathbb{A}^i_{n+r}\bigg)\sum_{j=0}^{r+h-i}p^{j}\,q^{(r+h-i-j)(m+s)}\, \mathbb{C}_{r+h-i}^j\, \mathbb{A}^j_{m+s}\nonumber\\
&
+\bigg(\sum_{i=0}^{h}p^{i}\,q^{(h-i)(m+s)}\, \mathbb{C}_h^i\, \mathbb{A}^i_{m+s}-\sum_{i=0}^sp^{i}\,q^{(s-i)(k+h)}\, \mathbb{C}_s^i\, \mathbb{A}^i_{k+h}\bigg)\nonumber\\
&\times\sum_{j=0}^{h+s-i}p^{j}\,q^{(h+s-i-j)(n+r)}\, \mathbb{C}_{h+s-i}^j\, \mathbb{A}^j_{n+r}\bigg](-1)^{i+j}
\lambda^{i+j-1}\bar{\mathcal{L}}_{m+n+k}^{s+r+h-i-j},
\end{align}
\begin{align}\label{Rqlim3}
\left[\mathcal{L}_m^s, \mathcal{L}_n^k, \mathcal{H}_r^{\alpha+\frac{1}{2}}\right]
&=\mathcal{K}(P,Q)\bigg[\bigg(\sum_{i=0}^{s}p^{i}\,q^{(s-i)(n+k)}\, \mathbb{C}_s^i\, \mathbb{A}^i_{n+k}-\sum_{i=0}^kp^{i}\,q^{(k-i)(m+s)}\,\mathbb{C}_k^i\,
\mathbb{A}^i_{m+s}\bigg)\nonumber\\&\times\sum_{j=0}^{s+k-i}p^{j}\,q^{(s+k-i-j)(r+\alpha)}\, \mathbb{C}_{s+k-i}^j\, \mathbb{A}^j_{r+\alpha}
+\bigg(\sum_{i=0}^{k}p^{i}\,q^{(k-i)(r+\alpha)}\, \mathbb{C}_k^i\, \mathbb{A}^i_{r+\alpha}\nonumber\\&-\sum_{i=0}^\alpha p^{i}\,q^{(\alpha-i)(n+k)}\, \mathbb{C}_\alpha^i\, \mathbb{A}^i_{n+k}\bigg)\sum_{j=0}^{k+\alpha-i}p^{j}\,q^{(k+\alpha-i-j)(m+s)} \,\mathbb{C}_{k+\alpha-i}^j\, \mathbb{A}^j_{m+s}\nonumber\\&
+\bigg(\sum_{i=0}^{\alpha}p^{i} \,q^{(\alpha-i)(m+s)}\,\mathbb{C}_{\alpha}^i\, \mathbb{A}^i_{m+s}-\sum_{i=0}^sp^{i}\,q^{(s-i)(r+\alpha)}\, \mathbb{C}_s^i \mathbb{A}^i_{r+\alpha}\bigg)\nonumber\\&\times\sum_{j=0}^{\alpha+s-i}p^{j}\,q^{(\alpha+s-i-j)(n+k)}\,
\mathbb{C}_{\alpha+s-i}^i\, \mathbb{A}^j_{n+k}\bigg](-1)^{i+j}\lambda^{i+j-1}\, \mathcal{H}_{m+n+r}^{s+k+\alpha-i-j+\frac{1}{2}},
\end{align}
\begin{align}\label{Rqlim4}
\left[\mathcal{L}_m^s, \mathcal{L}_n^k, \bar{\mathcal{H}}_r^{\alpha+\frac{1}{2}}\right]
&=\mathcal{K}(P,Q)\bigg[\bigg(\sum_{i=0}^{s}p^{i}\,q^{(s-i)(n+k)}\,\mathbb{C}_s^i\, \mathbb{A}^i_{n+k}-\sum_{i=0}^k p^{i}\,q^{(k-i)(m+s)}\,\mathbb{C}_k^i\,
\mathbb{A}^i_{m+s}\bigg)\nonumber\\&\times\sum_{j=0}^{s+k-i}p^{j}\,q^{(s+k-i-j)(r+\alpha)}\, \mathbb{C}_{s+k-i}^j\, \mathbb{A}^j_{r+\alpha}
+\bigg(\sum_{i=0}^{k}p^{i}\,q^{(k-i)(r+\alpha)} \, \mathbb{C}_k^i\, \mathbb{A}^i_{r+\alpha}\nonumber\\&-\sum_{i=0}^\alpha\,p^{i}\,q^{(\alpha-i)(n+k)}\, \mathbb{C}_\alpha^i\, \mathbb{A}^i_{n+k}\bigg)\sum_{j=0}^{k+\alpha-i}p^{j}\,q^{(k+\alpha-i-j)(m+s)} \, \mathbb{C}_{k+\alpha-i}^j\, \mathbb{A}^j_{m+s}\nonumber\\&
+\bigg(\sum_{i=0}^{\alpha}p^{i}\,q^{(\alpha-i)(m+s)}\,\mathbb{C}_{\alpha}^i\, \mathbb{A}^i_{m+s}-\sum_{i=0}^s\,p^{i}\,q^{(s-i)(r+\alpha)}\, \mathbb{C}_s^i \mathbb{A}^i_{r+\alpha}\bigg)\nonumber\\&\times
\sum_{j=0}^{\alpha+s-i}\,p^{j}\,q^{(\alpha+s-i-j)(n+k)}\mathbb{C}_{\alpha+s-i}^j\,\mathbb{A}^j_{n+k}]
(-1)^{i+j}\lambda^{i+j-1} \bar{\mathbb{H}}_{m+n+r}^{s+k+\alpha-i-j+\frac{1}{2}},
\end{align}
\begin{align}\label{Rqlim5}
\left[\mathcal{L}_m^k, \mathcal{H}_r^{\alpha+\frac{1}{2}}, \bar{\mathcal{H}}_s^{\beta+\frac{1}{2}}\right]
&=\mathcal{K}(P,Q)\bigg[\bigg(\sum_{i=0}^{k}p^{i}\,q^{(k-i)(r+\alpha)}\, \mathbb{C}_k^i\, \mathbb{A}^i_{r+\alpha}
-\sum_{i=0}^\alpha\,p^{i}\,q^{(\alpha-i)(m+k)}\,\mathbb{C}_\alpha^i \mathbb{A}^i_{m+k}\bigg)\nonumber\\&\times\sum_{j=0}^{k+\alpha-i}p^{j}\,q^{(k+\alpha-i-j)(s+\beta)}\,\mathbb{C}_{k+\alpha-i}^j
\,
\mathbb{A}^j_{s+\beta}
+\sum_{i=0}^{\alpha}\sum_{j=0}^{\alpha+\beta-i}p^{i}
\,q^{(\alpha-i)(s+\beta)}\nonumber\\&\times \mathbb{C}_\alpha^i\,	\mathbb{A}^i_{s+\beta}p^{j}q^{(\alpha+\beta-i-j)(m+k)}\,\mathbb{C}_{\alpha+\beta-i}^j\mathbb{A}^j_{m+k}\bigg](-1)^{i+j}\lambda^{i+j+1}
\mathcal{L}_{m+r+s}^{k+\alpha+\beta-i-j}\nonumber\\
&+\bigg[-\bigg(\sum_{i=0}^{k}p^{i}q^{(k-i)(r+\alpha)} \mathbb{C}_k^i \mathbb{A}^i_{r+\alpha}
-\sum_{i=0}^\alpha\,p^{i}\,q^{(\alpha-i)(m+k)} \mathbb{C}_\alpha^i \mathbb{A}^i_{m+k}\bigg)\nonumber\\&\times
\sum_{j=0}^{k+\alpha-i}p^{j}q^{(k+\alpha-i-j)(s+\beta)}\,\mathbb{C}_{k+\alpha-i}^j\,
\mathbb{A}^j_{s+\beta}
+\bigg(\sum_{i=0}^{\beta}p^{i}q^{(\beta-i)(r+\alpha)}\,\mathbb{C}_\beta^i\mathbb{A}^i_{r+\alpha}\nonumber\\
&-\sum_{i=0}^{\alpha}p^{i}q^{(\alpha-i)(s+\beta)}\,\mathbb{C}_\alpha^i\,\mathbb{A}^i_{s+\beta}\bigg)
\sum_{j=0}^{\alpha+\beta-i}p^{j}\,q^{(\alpha+\beta-i-j)(m+k)}\mathbb{C}_{\alpha+\beta-i}^j\,\mathbb{A}^j_{m+k}\nonumber\\&
-\bigg(\sum_{i=0}^\beta\,p^{i}\,q^{(\beta-i)(m+k)} \mathbb{C}_\beta^i \mathbb{A}^i_{m+k}
-\sum_{i=0}^{k}p^{i}\,q^{(k-i)(s+\beta)} \mathbb{C}_k^i
\mathbb{A}^i_{s+\beta}\bigg)\nonumber\\&\times
\sum_{j=0}^{k+\beta-i}p^{j}\,q^{(k+\beta-i-j)(r+\alpha)}\mathbb{C}_{k+\beta-i}^j\mathbb{A}^j_{r+\alpha}\bigg]
(-1)^{i+j} \lambda^{i+j-1}\bar{\mathcal{L}}_{m+r+s}^{k+\alpha+\beta-i-j}.
\end{align}
\begin{remark}
	Particular cases of super-high order Virasoro 3-algebra are deduced.
	\begin{enumerate}
		\item[(i)]
		Taking $\mathcal{R}(x)=\frac{x-1}{q-1}$ and $q\longrightarrow 1,$ we get the results given in \cite{CWZ}.
		\item[(ii)] Setting $\mathcal{R}(s,t)=(p-q)^{-1}(s-t)$, we deduce the $(p,q)$-super high-order Virasoro 3-algebra which satisfies the commutations relations \eqref{Rqlim1} to \eqref{Rqlim5} with $\mathcal{K}(P,Q)=1$ and \begin{eqnarray*}
			\mathbb{C}^{k}_{n} := \frac{[n]!_{p,q}}{[k]!_{p,q}[n-k]!_{p,q}},\quad  n\geq k,\quad \mathbb{A}_n^k:=
			\left\{\begin{array}{cc}
				[n]_{p,q}[n-1]_{p,q}
				\cdots[n-k+1]_{p,q},& k\leqslant n,\\
				0,                  &k>n.\end{array}\right. 
		\end{eqnarray*}
		\item[(iii)] Putting $\mathcal{R}(x,y)=\frac{x-y}{a\frac{x}{p}-b\frac{y}{q}},$ we derive the $(p,q)$-super high-order Virasoro 3-algebra which satisfies the commutations relations \eqref{Rqlim1} to \eqref{Rqlim5} with $\mathcal{K}(P,Q)=1$ and \begin{eqnarray*}
			\mathbb{C}^{k}_{n} := \frac{[n]^{a,b}_{p,q}!}{[k]^{a,b}_{p,q}![n-k]^{a,b}_{p,q}!},\quad  n\geq k,\quad \mathbb{A}_n^k:=
			\left\{\begin{array}{cc}
				[n]^{a,b}_{p,q}[n-1]^{a,b}_{p,q}
				\cdots[n-k+1]^{a,b}_{p,q},& k\leqslant n,\\
				0,                  &k>n.\end{array}\right. 
		\end{eqnarray*}
	\end{enumerate}
\end{remark}

To achieve our aims for this section,   we derive the $\mathcal{R}(p,q)$-super-$W_{\infty}$ 3-algebra by taking  the scaling limit $\lambda \rightarrow 0 .$  Then, by using the relation 
\eqref{Rqlim1} to \eqref{Rqlim5}, we obtain the $\mathcal{R}(p,q)$-super-$W_{\infty}$
3-algebra generated by the following commutations relations:
\begin{align}\label{Rq3alg1}
\left[\mathcal{L}_m^s,\mathcal{L}_n^r,\mathcal{L}_k^h\right]&=\mathcal{K}(P,Q)\bigg(q^{(s-1)(n+r)}[s]_{\mathcal{R}(p,q)}[n+r]_{\mathcal{R}(p,q)}-q^{(r-1)(m+s)}[r]_{\mathcal{R}(p,q)}\nonumber\\&\times[m+s]_{\mathcal{R}(p,q)}+q^{(r-1)(k+h)}[r]_{\mathcal{R}(p,q)}[k+h]_{\mathcal{R}(p,q)}-q^{(h-1)(n+r)}[h]_{\mathcal{R}(p,q)}\nonumber\\&\times[n+r]_{\mathcal{R}(p,q)}+q^{(h-1)(m+s)}[h]_{\mathcal{R}(p,q)}[m+s]_{\mathcal{R}(p,q)}-q^{(s-1)(k+h)}[s]_{\mathcal{R}(p,q)}\nonumber\\&\times[k+h]_{\mathcal{R}(p,q)}\bigg)
\mathcal{L}_{m+n+k}^{s+r+h-1},
\end{align}
\begin{align}\label{Rq3alg2}
\left[\mathcal{L}_m^s, \mathcal{L}_n^r, \bar {\mathcal{L}}_k^h\right]&=-\left[\mathcal{L}_m^s, \bar{\mathcal{L}}_k^h,  \mathcal{L}_n^r\right]=\left[\bar {\mathcal{L}}_k^h,  \mathcal{L}_m^s, \mathcal{L}_n^r\right]\nonumber\\
&=\mathcal{K}(P,Q)\bigg(q^{(s-1)(n+r)}[s]_{\mathcal{R}(p,q)}[n+r]_{\mathcal{R}(p,q)}-q^{(r-1)(m+s)}[r]_{\mathcal{R}(p,q)}\nonumber\\&\times[m+s]_{\mathcal{R}(p,q)}+q^{(r-1)(k+h)}[r]_{\mathcal{R}(p,q)}[k+h]_{\mathcal{R}(p,q)}-q^{(h-1)(n+r)}[h]_{\mathcal{R}(p,q)}\nonumber\\&\times[n+r]_{\mathcal{R}(p,q)}+q^{(h-1)(m+s)}[h]_{\mathcal{R}(p,q)}[m+s]_{\mathcal{R}(p,q)}-q^{(s-1)(k+h)}[s]_{\mathcal{R}(p,q)}\nonumber\\&\times[k+h]_{\mathcal{R}(p,q)}\bigg)\bar {\mathcal{L}}_{m+n+k}^{s+r+h-1},
\end{align}
\begin{small}
\begin{align}\label{Rq3alg3}
\left[\mathcal{L}_m^s, \mathcal{L}_n^k, \mathcal{H}_r^{\alpha+\frac{1}{2}}\right]&=-\left[\mathcal{L}_m^s,  \mathcal{H}_r^{\alpha+\frac{1}{2}}, \mathcal{L}_n^k\right]=
\left[\mathcal{H}_r^{\alpha+\frac{1}{2}}, \mathcal{L}_m^s, \mathcal{L}_n^k\right]\nonumber\\
&=\mathcal{K}(P,Q)\bigg(q^{(s-1)(n+k)}[s]_{\mathcal{R}(p,q)}[n+k]_{\mathcal{R}(p,q)}-q^{(k-1)(m+s)}[k]_{\mathcal{R}(p,q)}\nonumber\\&\times[m+s]_{\mathcal{R}(p,q)}+q^{(k-1)(r+\alpha)}[k]_{\mathcal{R}(p,q)}[r+\alpha]_{\mathcal{R}(p,q)}-q^{(\alpha-1)(n+k)}[\alpha]_{\mathcal{R}(p,q)}\nonumber\\&\times[n+k]_{\mathcal{R}(p,q)}+q^{(\alpha-1)(m+s)}[\alpha]_{\mathcal{R}(p,q)}[m+s]_{\mathcal{R}(p,q)}-q^{(s-1)(r+\alpha)}[s]_{\mathcal{R}(p,q)}\nonumber\\&\times[r+\alpha]_{\mathcal{R}(p,q)}\bigg)\mathcal{H}_{m+n+r}^{s+k+\alpha-1+\frac{1}{2}},
\end{align}
\begin{align}\label{Rq3alg4}
\left[\mathcal{L}_m^s, \mathcal{L}_n^k, \bar{\mathcal{H}}_r^{\alpha+\frac{1}{2}}\right]&=-\left[\mathcal{L}_m^s,  \bar{\mathcal{H}}_r^{\alpha+\frac{1}{2}}, \mathcal{L}_n^k\right]=
\left[\bar{\mathcal{H}}_r^{\alpha+\frac{1}{2}}, \mathcal{L}_m^s, \mathcal{L}_n^k\right]\nonumber\\
&=\mathcal{K}(P,Q)\bigg(q^{(s-1)(n+k)}[s]_{\mathcal{R}(p,q)}[n+k]_{\mathcal{R}(p,q)}-q^{(k-1)(m+s)}[k]_{\mathcal{R}(p,q)}\nonumber\\&\times[m+s]_{\mathcal{R}(p,q)}+q^{(k-1)(r+\alpha)}[k]_{\mathcal{R}(p,q)}[r+\alpha]_{\mathcal{R}(p,q)}-q^{(\alpha-1)(n+k)}[\alpha]_{\mathcal{R}(p,q)}\nonumber\\&\times[n+k]_{\mathcal{R}(p,q)}+q^{(\alpha-1)(m+s)}[\alpha]_{\mathcal{R}(p,q)}[m+s]_{\mathcal{R}(p,q)}-q^{(s-1)(r+\alpha)}[s]_{\mathcal{R}(p,q)}\nonumber\\&\times[r+\alpha]_{\mathcal{R}(p,q)}\bigg)\bar{\mathcal{H}}_{m+n+r}^{s+k+\alpha-1+\frac{1}{2}},
\end{align}
\begin{align}\label{Rq3alg5}
\left[\mathcal{L}_m^k, \mathcal{H}_r^{\alpha+\frac{1}{2}}, \bar{\mathcal{H}}_s^{\beta+\frac{1}{2}}\right]&=
\left[\mathcal{L}_m^k, \bar{\mathcal{H}}_s^{\beta+\frac{1}{2}}, \mathcal{H}_r^{\alpha+\frac{1}{2}}\right]
=\left[\mathcal{H}_r^{\alpha+\frac{1}{2}}, \bar{\mathcal{H}}_s^{\beta+\frac{1}{2}},\mathcal{L}_m^k\right]
=\left[\bar{\mathcal{H}}_s^{\beta+\frac{1}{2}}, \mathcal{H}_r^{\alpha+\frac{1}{2}}, \mathcal{L}_m^k\right]\nonumber\\
&=-\left[\mathcal{H}_r^{\alpha+\frac{1}{2}}, \mathcal{L}_m^k, \bar{\mathcal{H}}_s^{\beta+\frac{1}{2}}\right]
=-\left[\bar{\mathcal{H}}_s^{\beta+\frac{1}{2}}, \mathcal{L}_m^k, \mathcal{H}_r^{\alpha+\frac{1}{2}}\right]\nonumber\\
&=\mathcal{K}(P,Q)\bigg(-q^{(k-1)(r+\alpha)}[k]_{\mathcal{R}(p,q)}[r+\alpha]_{\mathcal{R}(p,q)}+q^{(\alpha-1)(m+k)}[\alpha]_{\mathcal{R}(p,q)}\nonumber\\&\times[m+k]_{\mathcal{R}(p,q)}+q^{(\beta-1)(r+\alpha)}[\beta]_{\mathcal{R}(p,q)}[r+\alpha]_{\mathcal{R}(p,q)}-q^{(\alpha-1)(s+\beta)}[\alpha]_{\mathcal{R}(p,q)}\nonumber\\&\times[s+\beta]_{\mathcal{R}(p,q)}-q^{(\beta-1)(m+k)}[\beta]_{\mathcal{R}(p,q)}[m+k]_{\mathcal{R}(p,q)}+q^{(k-1)(s+\beta)}[k]_{\mathcal{R}(p,q)}\nonumber\\&\times[s+\beta]_{\mathcal{R}(p,q)}\bigg)\bar
{\mathcal{L}}_{m+r+s}^{i+\alpha+\beta-1},
\end{align}
\end{small}
with all other 3-brackets vanishing. 

Note
that the 3-algebraic relations \eqref{Rq3alg1},\eqref{Rq3alg2} and \eqref{Rq3alg3} generated the $\mathcal{R}(p,q)$-deformation of the $w_{\infty}$ 3-algebra introduced in \cite{CKJ}. It is
known that this $W_{\infty}$ 3-algebra satisfies the usualy FI
condition. As to the quantum super $W_{\infty}$ $3$-algebra \eqref{Rq3alg1},
it should satisfy the generalized FI condition due to the involution
of fermionic generators. We find  that the quantum super $W_{\infty}$
$3$-algebra \eqref{Rq3alg1} satisfies the following generalized FI
condition:
\begin{eqnarray}\label{eq:FI}
[A, B, [C, D, E]]&=&[[A, B, C], D, E]+(-1)^{(|A|+|B|)|C|}[C, [A, B, D], E]\nonumber\\
&+&(-1)^{(|A|+|B|)(|C|+|D|)}[C, D, [A, B, E]].
\end{eqnarray}
When the generators in (\ref{eq:FI}) are bosonic, (\ref{eq:FI}) reduces to the usually FI condition of 3-algebra.
\begin{remark}
	Particular super $w_{\infty}$ 3-algebra from quantum algebras are deduced as:
	\begin{enumerate}
		\item[(i)] Taking $\mathcal{R}(x,y)=\frac{x-y}{p-q},$ we deduce the $(p,q)$-deformed super $w_{\infty}$ 3-algebra as follows: 
		\begin{align*}
		\left[\mathcal{L}_m^s,\mathcal{L}_n^r,\mathcal{L}_k^h\right]&=\bigg(q^{(s-1)(n+r)}[s]_{p,q}[n+r]_{p,q}-q^{(r-1)(m+s)}[r]_{\mathcal{R}(p,q)}[m+s]_{p,q}\nonumber\\&+q^{(r-1)(k+h)}[r]_{p,q}[k+h]_{p,q}-q^{(h-1)(n+r)}[h]_{p,q}[n+r]_{p,q}\nonumber\\&+q^{(h-1)(m+s)}[h]_{p,q}[m+s]_{p,q}-q^{(s-1)(k+h)}[s]_{p,q}[k+h]_{p,q}\bigg)
		\mathcal{L}_{m+n+k}^{s+r+h-1},
		\end{align*}
		\begin{align*}
		\left[\mathcal{L}_m^s, \mathcal{L}_n^r, \bar {\mathcal{L}}_k^h\right]&=-\left[\mathcal{L}_m^s, \bar{\mathcal{L}}_k^h,  \mathcal{L}_n^r\right]=\left[\bar {\mathcal{L}}_k^h,  \mathcal{L}_m^s, \mathcal{L}_n^r\right]\nonumber\\
		&=\bigg(q^{(s-1)(n+r)}[s]_{p,q}[n+r]_{p,q}-q^{(r-1)(m+s)}[r]_{p,q}[m+s]_{p,q}\nonumber\\&+q^{(r-1)(k+h)}[r]_{p,q}[k+h]_{p,q}-q^{(h-1)(n+r)}[h]_{p,q}[n+r]_{p,q}\nonumber\\&+q^{(h-1)(m+s)}[h]_{p,q}[m+s]_{p,q}-q^{(s-1)(k+h)}[s]_{p,q}[k+h]_{p,q}\bigg)\bar {\mathcal{L}}_{m+n+k}^{s+r+h-1},
		\end{align*}
		\begin{align*}
		\left[\mathcal{L}_m^s, \mathcal{L}_n^k, \mathcal{H}_r^{\alpha+\frac{1}{2}}\right]&=-\left[\mathcal{L}_m^s,  \mathcal{H}_r^{\alpha+\frac{1}{2}}, \mathcal{L}_n^k\right]=
		\left[\mathcal{H}_r^{\alpha+\frac{1}{2}}, \mathcal{L}_m^s, \mathcal{L}_n^k\right]\nonumber\\
		&=\bigg(q^{(s-1)(n+k)}[s]_{p,q}[n+k]_{p,q}-q^{(k-1)(m+s)}[k]_{p,q}[m+s]_{p,q}\nonumber\\&+q^{(k-1)(r+\alpha)}[k]_{p,q}[r+\alpha]_{p,q}-q^{(\alpha-1)(n+k)}[\alpha]_{p,q}[n+k]_{p,q}\nonumber\\&+q^{(\alpha-1)(m+s)}[\alpha]_{p,q}[m+s]_{p,q}-q^{(s-1)(r+\alpha)}[s]_{p,q}[r+\alpha]_{p,q}\bigg)\mathcal{H}_{m+n+r}^{s+k+\alpha-1+\frac{1}{2}},
		\end{align*}
		\begin{align*}
		\left[\mathcal{L}_m^s, \mathcal{L}_n^k, \bar{\mathcal{H}}_r^{\alpha+\frac{1}{2}}\right]&=-\left[\mathcal{L}_m^s,  \bar{\mathcal{H}}_r^{\alpha+\frac{1}{2}}, \mathcal{L}_n^k\right]=
		\left[\bar{\mathcal{H}}_r^{\alpha+\frac{1}{2}}, \mathcal{L}_m^s, \mathcal{L}_n^k\right]\nonumber\\
		&=\bigg(q^{(s-1)(n+k)}[s]_{p,q}[n+k]_{p,q}-q^{(k-1)(m+s)}[k]_{p,q}[m+s]_{p,q}\nonumber\\&+q^{(k-1)(r+\alpha)}[k]_{p,q}[r+\alpha]_{p,q}-q^{(\alpha-1)(n+k)}[\alpha]_{p,q}[n+k]_{p,q}\nonumber\\&+q^{(\alpha-1)(m+s)}[\alpha]_{p,q}[m+s]_{p,q}-q^{(s-1)(r+\alpha)}[s]_{p,q}[r+\alpha]_{p,q}\bigg)\bar{\mathcal{H}}_{m+n+r}^{s+k+\alpha-1+\frac{1}{2}},
		\end{align*}
		\begin{align*}
		\left[\mathcal{L}_m^k, \mathcal{H}_r^{\alpha+\frac{1}{2}}, \bar{\mathcal{H}}_s^{\beta+\frac{1}{2}}\right]&=
		\left[\mathcal{L}_m^k, \bar{\mathcal{H}}_s^{\beta+\frac{1}{2}}, \mathcal{H}_r^{\alpha+\frac{1}{2}}\right]
		=\left[\mathcal{H}_r^{\alpha+\frac{1}{2}}, \bar{\mathcal{H}}_s^{\beta+\frac{1}{2}},\mathcal{L}_m^k\right]
		=\left[\bar{\mathcal{H}}_s^{\beta+\frac{1}{2}}, \mathcal{H}_r^{\alpha+\frac{1}{2}}, \mathcal{L}_m^k\right]\nonumber\\
		&=-\left[\mathcal{H}_r^{\alpha+\frac{1}{2}}, \mathcal{L}_m^k, \bar{\mathcal{H}}_s^{\beta+\frac{1}{2}}\right]
		=-\left[\bar{\mathcal{H}}_s^{\beta+\frac{1}{2}}, \mathcal{L}_m^k, \mathcal{H}_r^{\alpha+\frac{1}{2}}\right]\nonumber\\
		&=\bigg(-q^{(k-1)(r+\alpha)}[k]_{p,q}[r+\alpha]_{p,q}+q^{(\alpha-1)(m+k)}[\alpha]_{p,q}[m+k]_{p,q}\nonumber\\&+q^{(\beta-1)(r+\alpha)}[\beta]_{p,q}[r+\alpha]_{p,q}-q^{(\alpha-1)(s+\beta)}[\alpha]_{p,q}[s+\beta]_{p,q}\nonumber\\&-q^{(\beta-1)(m+k)}[\beta]_{p,q}[m+k]_{p,q}+q^{(k-1)(s+\beta)}[k]_{p,q}[s+\beta]_{p,q}\bigg)\bar
		{\mathcal{L}}_{m+r+s}^{i+\alpha+\beta-1},
		\end{align*}
		with all other 3-brackets vanishing.
	\end{enumerate}
\end{remark}
\section{$\mathcal{R}(p,q)$-super $n$-algebra}
Here, we investigate the super $n$-algebra from the generalized quantum algebra\cite{HB}. 

Let $A_i,\quad i\in\{1,2,\cdots,n\}$ be arbitrary associative operators. Then, the $\mathcal{R}(p,q)$-super
multibracket of order $n$ is defined by:
\begin{align}\label{eq1}
\left[A_1,A_2,\cdots,A_n\right]&=\frac{1}{2^{\alpha}}\bigg(\frac{[-2\sum_{l=1}^{n}m_{l}]_{\mathcal{R}(p,q)}}{[-\sum_{l=1}^{n}m_{l}]_{\mathcal{R}(p,q)}}\bigg)^{\alpha}\epsilon_{1 2 \cdots n}^{i_1 i_2 \cdots i_n}\nonumber\\&\times\big(-1\big)^{\sum_{k=1}^{n-1}|A_k|(\sum_{l=k+1,i_l< i_k}^{n}|A_{i_l}|)}\,A_{i_1}\,A_{i_2}\cdots A_{i_n},
\end{align}
where $\alpha=\frac{1+(-1)^n }{ 2},$    
 $\epsilon^{i_1i_2\cdots i_n}_{12\cdots n}$ is the L\' evi-Civita symbol defined by:
\begin{eqnarray}\label{eq2}
\epsilon^{j_1 \cdots j_p}_{i_1 \cdots i_p}:= det\left( \begin{array} {ccc}
\delta^{j_1}_{i_1} &\cdots&  \delta^{j_1}_{i_p}   \\ 
\vdots && \vdots \\
\delta^{j_p}_{i_1} & \cdots& \delta^{j_p}_{i_p}
\end {array} \right) ,
\end{eqnarray}
and the symbol $|A|$ is to be understood as the parity of $A$.

For the bosonic operators, the relation \eqref{eq1} is reduced to the $n$-commutator:
\begin{eqnarray}\label{eq3}
\left[A_1,A_2,\cdots,A_n\right]=\frac{1}{2^{\alpha}}\bigg(\frac{[-2\sum_{l=1}^{n}m_{l}]_{\mathcal{R}(p,q)}}{[-\sum_{l=1}^{n}m_{l}]_{\mathcal{R}(p,q)}}\bigg)^{\alpha}\epsilon_{1 2 \cdots n}^{i_1 i_2 \cdots i_n}\,A_{i_1}\,A_{i_2}\cdots A_{i_n}.
\end{eqnarray}
The first $\mathcal{R}(p,q)$-super multibrackets are deduced as follows:
\begin{enumerate}
	\item[(i)] For $n=2,$ we have:
\begin{eqnarray}\label{eq4}
\left[A_1,A_2\right]=\frac{1}{2}\bigg(\frac{[-2\big(m_{1}+m_{2}\big)]_{\mathcal{R}(p,q)}}{[-\big(m_{1}+m_{2}\big)]_{\mathcal{R}(p,q)}}\bigg)\bigg(A_1\,A_2-(-1)^{|A_2||A_1|}A_2\,A_1\bigg).
\end{eqnarray}
Note that, the relation \eqref{eq4} can be  considered as the $\mathcal{R}(p,q)$-Lie super bracket.
\item[(ii)] Taking $n=3,$ we obtain:
\begin{align}\label{eq5}
\left[A_1,A_2,A_3\right]&=A_1A_2A_3-(-1)^{|A_2||A_3|}A_1A_3A_2-(-1)^{|A_1||A_2|}A_2A_1A_3\nonumber\\
&+(-1)^{|A_1|\big(|A_2|+|A_3|\big)}A_2A_3A_1+(-1)^{|A_3|\big(|A_1|+|A_2|\big)}A_3A_1A_2\nonumber\\
&-(-1)^{|A_3|\big(|A_1|+|A_2|\big)+|A_1||A_2|}A_3A_2A_1\nonumber\\
&=A_1[A_2,A_3]-(-1)^{|A_1||A_2|}A_2[A_1,A_3]+(-1)^{|A_3|\big(|A_1|+|A_2|\big)}A_3[A_1,A_2].
\end{align}
\item[(iii)] For $n=4$, we get:
\begin{small}
\begin{align}\label{eq6}
\left[A_1,A_2,A_3,A_4\right]&=\mathcal{A}(p,q)\bigg(A_1[A_2,A_3,A_4]-(-1)^{|A_1||A_2|}A_2[A_1,A_3,A_4]\nonumber\\
&+(-1)^{|A_3|\big(|A_1|+|A_2|\big)}A_3[A_1,A_2,A_4]-(-1)^{|A_4|\big(|A_3|+|A_2|+|A_1|\big)}A_4[A_1,A_2,A_3]\bigg)\nonumber\\
&=\mathcal{A}(p,q)\bigg([A_1, A_2][A_3,A_4]+(-1)^{(|A_4|+|A_3)(||A_2|+|A_1|)}[A_3, A_4][A_1,A_2]\nonumber\\&-(-1)^{|A_3||A_2|}[A_1, A_3][A_2,A_4]-(-1)^{|A_4|\big(|A_1|+|A_3|\big)+|A_2||A_1|}[A_2, A_4][A_1,A_3]\nonumber\\&+(-1)^{|A_4|\big(|A_2|+|A_3|\big)}[A_1, A_4][A_2,A_3]+(-1)^{|A_1|\big(|A_2|+|A_3|\big)}[A_2, A_3][A_1,A_4]\bigg),
\end{align}
\end{small}
where $\mathcal{A}(p,q)=\frac{[-2\sum_{l=1}^{4}m_l]_{\mathcal{R}(p,q)}}{2[-\sum_{l=1}^{4}m_l]_{\mathcal{R}(p,q)}}.$
\end{enumerate}
From the relation \eqref{eq1}, we have the  deformed skewsymmetry
\begin{align}\label{eq7}
\left[A_{\sigma_1},A_{\sigma_2},\cdots,A_{\sigma_n}\right]&=\frac{1}{2^{\alpha}}\bigg(\frac{[-2\sum_{l=1}^{n}m_{l}]_{\mathcal{R}(p,q)}}{[-\sum_{l=1}^{n}m_{l}]_{\mathcal{R}(p,q)}}\bigg)^{\alpha}\,\epsilon_{1 2 \cdots n}^{i_1 i_2 \cdots i_n}\nonumber\\&\times\big(-1\big)^{\sum_{k=1}^{n-1}|A_{\sigma_k}|(\sum_{l=k+1,i_l< i_k}^{n}|A_{\sigma_l}|)}\,\left[A_{1},A_{2},\cdots,A_{n}\right].
\end{align}
When $n$ is even, the $\mathcal{R}(p,q)$-super multibracket \eqref{eq1} satisfies the generalized super Jacobi identity (GSJI)\cite{HW}:
\begin{eqnarray}\label{eq8}
\epsilon_{1 2 \cdots n}^{i_1 i_2 \cdots i_{2n-1}}(-1)^{\sum_{k=1}^{2n-1}|A_k|(\sum_{l=k+1,i_l< i_k}^{2n-1}|A_{i_l}|)}\left[\left[A_{i_1},\,A_{i_2}\cdots A_{i_n}\right],A_{i_{n+1}},A_{i_{n+2}},\cdots,A_{i_{2n-1}}\right]=0.
\end{eqnarray}
For the generators $X_i$, the generalized super Lie algebra of order $n=2k$ is defined by an expression of the form
that satisfies the GSJI \eqref{eq8}
\begin{eqnarray}\label{eq9}
\left[X_{i_1},X_{i_2},\cdots,X_{i_n}\right]=C^{j}_{i_1i_2\cdots i_{2k}}\,X_j,
\end{eqnarray}
where the multibracket is defined by the relation \eqref{eq1} and $C^{j}_{i_1i_2\cdots i_{2k}}$
are the structure constants.

For the odd $n$-brackets \eqref{eq2} built from associative operator products, it is known that they satisfy the generalized
Bremner identity (GBI)\cite{Br1,Br2,CFJMZ,DF}:
\begin{align}\label{eq10}
&\epsilon_{1  \cdots 3n-3}^{i_1 \cdots i_{3n-3}}\left[\left[B,A_{i_1},\cdots A_{i_{n-1}}\right],\left[A_{i_{n}},\cdots,A_{i_{2n-1}}\right],A_{i_{2n}},\cdots,A_{i_{3n-3}}\right]\nonumber\\&=\epsilon_{1  \cdots 3n-3}^{i_1 \cdots i_{3n-3}}\left[\left[B,\left[A_{i_1},\cdots,A_{i_{n}}\right],A_{i_{n+1}},\cdots,A_{i_{2n-2}}\right],A_{i_{2n-1}},\cdots,A_{i_{3n-3}}\right].
\end{align}
However for the case of the super odd $n$-brackets, the corresponding generalized super Bremner identity (GSBI), to
our best knowledge, has not been reported so far in the existing literature. Let us deal with it below.

According to the definition of n bracket \eqref{eq1}, we have:
\begin{align}\label{eq11}
\left[B,A_{1},A_{2},\cdots A_{n-1}\right]&=\frac{1}{2^{\alpha}}\bigg(\frac{[-2\sum_{l=1}^{n}m_{l}]_{\mathcal{R}(p,q)}}{[-\sum_{l=1}^{n}m_{l}]_{\mathcal{R}(p,q)}}\bigg)^{\alpha}\sum_{j=0}^{n-1}(-1)^{j}\epsilon_{12  \cdots (n-1)}^{i_1i_2 \cdots i_{n-1}}\nonumber\\&\times(-1)^{|p_1|}A_{i_1},A_{i_2},\cdots,A_{i_{j}}BA_{i_{j+1}}\cdots A_{i_{n-1}},
\end{align}
\begin{align}\label{eq12}
\left[B,A_{1},A_{2},\cdots A_{n},Z\right]&=\frac{1}{2^{\alpha}}\bigg(\frac{[-2\sum_{l=1}^{n}m_{l}]_{\mathcal{R}(p,q)}}{[-\sum_{l=1}^{n}m_{l}]_{\mathcal{R}(p,q)}}\bigg)^{\alpha}\bigg(\sum_{j=0}^{n}(-1)^{j}\sum_{k=0}^{n-j}(-1)^k\epsilon_{12  \cdots n}^{i_1i_2 \cdots i_{n}}\nonumber\\&\times(-1)^{|p_2|} A_{i_1},\cdots,A_{i_{k}}BA_{i_{k+1}}\cdots A_{i_{n-j}}ZA_{i_{n-j+1}}\cdots A_{i_{n}}\nonumber\\
&-\sum_{j=0}^{n}(-1)^{j}\sum_{k=0}^{n-j}(-1)^k\epsilon_{12  \cdots n}^{i_1i_2 \cdots i_{n}}\nonumber\\&\times(-1)^{|p_3|} A_{i_1},\cdots,A_{i_{k}}ZA_{i_{k+1}}\cdots A_{i_{n-j}}BA_{i_{n-j+1}}\cdots A_{i_{n}}\bigg),
\end{align}
for any fixed $B$ and $Z$, where
\begin{eqnarray}\label{eq13}
|p_1|=\sum_{k=2}^{n-2}|A_{i_k}|\bigg(\sum_{l=k+1,i_l<i_k}^{n-1}|A_{i_l}|\bigg)+|B|\big(|A_{i_1}|+\cdots+|A_{i_j}|\big),
\end{eqnarray}
\begin{eqnarray}\label{eq14}
|p_2|=\sum_{s=1}^{n-1}|A_{i_s}|\bigg(\sum_{l=s+1,i_l<i_s}^{n}|A_{i_l}|\bigg)+|B|\big(|A_{i_1}|+\cdots+|A_{i_k}|\big)+|Z|\big(|A_{i_{n-j+1}}|+\cdots+|A_{i_n}|\big),
\end{eqnarray}
and
\begin{small}
\begin{eqnarray}\label{eq15}
|p_3|=\sum_{s=1}^{n-1}|A_{i_s}|\bigg(\sum_{l=s+1,i_l<i_s}^{n}|A_{i_l}|\bigg)+|B|\big(|A_{i_1}|+\cdots+|A_{i_{n-j}}|\big)+|Z|\big(|B|+|A_{i_{k+1}}|+\cdots+|A_{i_n}|\big).
\end{eqnarray}
\end{small}
By using the relations \eqref{eq11} and \eqref{eq12}, we can show (see Appendix ) that when $n$ is odd, the $\mathcal{R}(p,q)$-super $n$-bracket \eqref{eq1} satisfies
the following GSBI:
\begin{align}\label{eq16}
&\epsilon_{1  \cdots 3n-3}^{i_1 \cdots i_{3n-3}}(-1)^{\sum_{k=1}^{3n-4}|A_{i_k}|(\sum_{l=k+1,i_l< i_k}^{3n-3}|A_{i_l}|)}\nonumber\\&\times\left[\left[B,A_{i_1},\cdots A_{i_{n-1}}\right],\left[A_{i_{n}},\cdots,A_{i_{2n-1}}\right],A_{i_{2n}},\cdots,A_{i_{3n-3}}\right]\nonumber\\&=\epsilon_{1  \cdots 3n-3}^{i_1 \cdots i_{3n-3}}(-1)^{\sum_{k=1}^{3n-4}|A_{i_k}|(\sum_{l=k+1,i_l< i_k}^{3n-3}|A_{i_l}|)}\nonumber\\&\times\left[\left[B,\left[A_{i_1},\cdots,A_{i_{n}}\right],A_{i_{n+1}},\cdots,A_{i_{2n-2}}\right],A_{i_{2n-1}},\cdots,A_{i_{3n-3}}\right].
\end{align}
For the case of the bosonic operators, the relation \eqref{eq16} is reduced to the
GBI given by \eqref{eq10}.
\section{$\mathcal{R}(p,q)$-super $W_{1+\infty}$ $n$-algebra in the Supersymmetric Landau Problem}
We construct the super $W_{1+\infty}$ $n$-algebra in the framework of the quantum algebra \cite{HB}.  

Let us consider a supersymmetric Landau problem,
where the Hamiltonian is given by \cite{H1}:
\begin{eqnarray}\label{eq17}
H=\frac{1}{2m}\bigg(P^2_i+C_{\alpha\beta}P_{\alpha}P_{\beta}\bigg),
\end{eqnarray}
where $C_{\alpha\beta}$ denotes the $\alpha-th$ row and $\beta-th$ column element
in a charge conjugation matrix
\begin{equation}\label{eq18}
C=i\,\sigma_{2}=\left( \begin{array} {cc}  0 & 1  \\
	-1   & 0
	\end {array} \right),
\end{equation}
the covariant momenta are given by:
\begin{equation}\label{eq19}
P_i=-i\big(\partial_i+i\big(\tilde{B}/2\big)\epsilon_{ij}x_j\big)\quad\mbox{and}\quad P_{\alpha}=i\big(\partial_{\alpha}-\big(\tilde{B}/2\big)\epsilon^2_{\alpha\beta}\theta_{\beta}\big),
\end{equation}
where $i=\sqrt{-1},\, \epsilon_{11}=\epsilon_{22}=0,\,\epsilon_{12}=-\epsilon_{21}=1,\,\partial_i=\partial/\partial x_i,\,i\in\{1,2\},\,\partial_{\alpha}=\partial/\partial \theta_{\alpha},\,\alpha\in\{1,2\},$ $\tilde{B}$ is a constant magnetic
field, $x_i$ and $\theta_{\alpha}$ are the bosonic and fermionic coordinates,
respectively.
\begin{align}\label{eq20}
\,z&=\frac{1}{2\,l_{\tilde{B}}}\big(x+iy\big),\quad z^{\star}=\frac{1}{2\,l_{\tilde{B}}}\big(x-iy\big)\nonumber\\
\,\partial&=l_{\tilde{B}}\big(\partial_x-i\partial_y\big),\quad \partial^{\star}=l_{\tilde{B}}\big(\partial_x+i\partial_y\big),
\end{align}
\begin{align}\label{eq21}
\,\theta&=\frac{1}{\sqrt{2}l_{\tilde{B}}}\,\theta_1,\quad \theta^{\star}=\frac{1}{\sqrt{2}l_{\tilde{B}}}\,\theta_2\nonumber\\
\,\partial_{\theta}&=\sqrt{2}l_{\tilde{B}}\,\partial_{\theta_1},\quad \partial_{\theta}=\sqrt{2}l_{\tilde{B}}\,\partial_{\theta_2}, 
\end{align}
where $l_{\tilde{B}}=1/\sqrt{\tilde{B}}$ is the magnetic length.

In the symmetric gauge, we may define the bosonic
creation and annihilation operators:
\begin{align}\label{eq22}
a=-\frac{i}{\sqrt{2}}\big(z+\partial^{\star}\big),\quad a^{\dagger}=\frac{i}{\sqrt{2}}\big(z^{\star}-\partial\big),
\end{align}
\begin{align}\label{eq23}
b=\frac{1}{\sqrt{2}}\big(z^{\star}+\partial\big),\quad b^{\dagger}=\frac{1}{\sqrt{2}}\big(z-\partial^{\star}\big),
\end{align}
and the fermionic creation and annihilation operators
\begin{align}\label{eq24}
\alpha=-\frac{i}{\sqrt{2}}\big(\theta-\partial^{\star}_{\theta}\big),\quad \alpha^{\dagger}=-\frac{i}{\sqrt{2}}\big(\theta^{\star}-\partial_{\theta}\big),
\end{align}
\begin{align}\label{eq25}
\beta=\frac{1}{\sqrt{2}}\big(\theta^{\star}+\partial_{\theta}\big),\quad \beta^{\dagger}=\frac{1}{\sqrt{2}}\big(\theta+\partial^{\star}_{\theta}\big).
\end{align}
They satisfy $[a,\,a^{\dagger}]=[b,\,b^{\dagger}]=1$ and $\{\alpha,\,\alpha^{\dagger}\}=\{\beta,\,\beta^{\dagger}\}=1.$ Other (anti) commutators are zeros.

In terms of the creation and annihilation operators,
the Hamiltonian \eqref{eq17} can be rewritten as\cite{H1}:
\begin{eqnarray}\label{eq26}
H=\frac{\tilde{B}}{m}\big(a^{\dagger}a+\alpha^{\dagger}\alpha\big).
\end{eqnarray}
We may also express the angular momentum $L_{\perp}$ as:
\begin{align}\label{eq27}
L_{\perp}&=(\sigma_2)_{ij}x_i\partial_j+\frac{1}{2}(\sigma_3)_{\alpha\beta}\theta_{\alpha}\partial_{\beta}\nonumber\\
&=\big(b^{\dagger}b+\frac{1}{2}\beta^{\dagger}\beta
\big)-\big(a^{\dagger}a+\frac{1}{2}\alpha^{\dagger}\alpha\big),
\end{align}
where $\sigma_2$ and $\sigma_3$ are the Pauli matrices. It commutes with
the Hamiltonian \eqref{eq26}.

Let us consider the  deformed operators
\begin{align}\label{eq28}
W^{B}_{m,r}&=b^{{\dagger}^{m+r-1}}b^{r-1},\quad \bar{W}^{B}_{m,r}=b^{{\dagger}^{m+r-1}}b^{r-1}\beta^{\dagger}\beta,
\end{align}
\begin{align}\label{eq29}
W^{F}_{m,r}&=b^{{\dagger}^{m+r-1}}b^{r-1}\beta,\quad \bar{W}^{F}_{m,r}=b^{{\dagger}^{m+r-1}}b^{r-1}\beta^{\dagger},
\end{align}
where $m\geq 1$ and $m+r\geq 1,$ the superindices B and F
denote the corresponding operators being the bosonic and
fermionic ones, respectively. It can be  easily seen
that the operators \eqref{eq28} and \eqref{eq29} commute with the Hamiltonian
\eqref{eq26}. It indicates that they are indeed conserved
operators.

By calculating the (anti) commutation relations, we
find that the  conserved operators \eqref{eq28} and \eqref{eq29} satisfy the $\mathcal{R}(p,q)$-super $W_{1+\infty}$ algebra:
\begin{align}\label{eq30}
\left[W^{B}_{m_1,r_1},\,W^{B}_{m_2,r_2}\right]&=f^{m_1,r_1}_{m_2,r_2}\,W^{B}_{\bar{m},\bar{r}},\quad \left[W^{B}_{m_1,r_1},\,\bar{W}^{B}_{m_2,r_2}\right]=f^{m_1,r_1}_{m_2,r_2}\,\bar{W}^{B}_{\bar{m},\bar{r}},
\end{align}
\begin{align}\label{eq31}
\left[W^{B}_{m_1,r_1},\,W^{F}_{m_2,r_2}\right]&=f^{m_1,r_1}_{m_2,r_2}\,W^{F}_{\bar{m},\bar{r}},\quad \left[W^{B}_{m_1,r_1},\,\bar{W}^{F}_{m_2,r_2}\right]=f^{m_1,r_1}_{m_2,r_2}\,\bar{W}^{F}_{\bar{m},\bar{r}},
\end{align}
\begin{align}\label{eq32}
\left[\bar{W}^{B}_{m_1,r_1},\,\bar{W}^{B}_{m_2,r_2}\right]&=f^{m_1,r_1}_{m_2,r_2}\,\bar{W}^{B}_{\bar{m},\bar{r}},\quad \left[W^{F}_{m_1,r_1},\,W^{F}_{m_2,r_2}\right]=\left[\bar{W}^{F}_{m_1,r_1},\,\bar{W}^{F}_{m_2,r_2}\right]=0,
\end{align}

\begin{align}\label{eq33}
\left[\bar{W}^{B}_{m_1,r_1},\,W^{F}_{m_2,r_2}\right]&=-K(P,Q)\,p^{\alpha_1}\sum_{\alpha_1=0}^{r_2-1}q^{(r_2-1-\alpha_1)(m_1+r_1-1)}\mathbb{C}^{\alpha_1}_{r_2-1}\,\mathbb{A}^{\alpha_1}_{m_1+r_1-1}\,W^{F}_{\bar{m},\bar{r}},\end{align}
\begin{align}\label{eq34}
 \left[\bar{W}^{B}_{m_1,r_1},\,\bar{W}^{F}_{m_2,r_2}\right]&=K(P,Q)\,p^{\alpha_1}\sum_{\alpha_1=0}^{r_1-1}q^{(r_1-1-\alpha_1)(m_2+r_2-1)}\,\mathbb{C}^{\alpha_1}_{r_1-1}\,\mathbb{A}^{\alpha_1}_{m_2+r_2-1}\,\bar{W}^{F}_{\bar{m},\bar{r}},
\end{align}
\begin{align}\label{eq35}
\left[W^{F}_{m_1,r_1},\,\bar{W}^{F}_{m_2,r_2}\right]&=K(P,Q)\,p^{\alpha_1}\sum_{\alpha_1=0}^{r_1-1}q^{(r_1-1-\alpha_1)(m_2+r_2-1)}\,\mathbb{C}^{\alpha_1}_{r_1-1}\,\mathbb{A}^{\alpha_1}_{m_2+r_2-1}\,W^{B}_{\bar{m},\bar{r}}\nonumber\\&-K(P,Q)\,f^{m_1,r_1}_{m_2,r_2}\,\bar{W}^{B}_{\bar{m},\bar{r}},
\end{align}
where $\bar{m}=m_1+m_2,$ $\bar{r}=r_1+r_2-1-\alpha_1$ and 
\begin{align}\label{eq36}
f^{m_1,r_1}_{m_2,r_2}&=K(P,Q)\,p^{\alpha_1}\bigg(\sum_{\alpha_1=0}^{r_1-1}q^{(r_1-1-\alpha_1)(m_2+r_2-1)}\mathbb{C}^{\alpha_1}_{r_1-1}\,\mathbb{A}^{\alpha_1}_{m_2+r_2-1}\nonumber\\&-\sum_{\alpha_1=0}^{r_1-1}q^{(r_2-1-\alpha_1)(m_1+r_1-1)}\,\mathbb{C}^{\alpha_1}_{r_2-1}\,\mathbb{A}^{\alpha_1}_{m_1+r_1-1}\bigg).
\end{align}
 
In the bosonic
limit, the relations \eqref{eq30} to \eqref{eq35} are reduced to the $\mathcal{R}(p,q)$-$W_{1+\infty}$ algebra. Note that, by taking $\mathcal{R}(p,q)=1,$ we recover the $W_{1+\infty}$ algebra described in \cite{CTZ1}.

For the $\mathcal{R}(p,q)$-operators \eqref{eq28} and \eqref{eq29}, the commutation
relations with the angular momentum operators are given by:
\begin{align}\label{eq37}
\,\left[L_{\perp},\,W^{B}_{m,r}\right]&=[m]_{\mathcal{R}(p,q)}\,W^{B}_{m,r}\nonumber\\
\,\left[L_{\perp},\,\bar{W}^{B}_{m,r}\right]&=[m]_{\mathcal{R}(p,q)}\,\bar{W}^{B}_{m,r}\nonumber\\
\,\left[L_{\perp},\,W^{F}_{m,r}\right]&=[m-\frac{1}{2}]_{\mathcal{R}(p,q)}\,W^{F}_{m,r}\nonumber\\
\,\left[L_{\perp},\,\bar{W}^{F}_{m,r}\right]&=[m+\frac{1}{2}]_{\mathcal{R}(p,q)}\,\bar{W}^{F}_{m,r}.
\end{align}

Now we turn us to the $\mathcal{R}(p,q)$-super 3-bracket \eqref{eq5}. By using the relation 
\eqref{eq37}, we can deduce the $\mathcal{R}(p,q)$-super 3-brackets as follows:
\begin{align}\label{eq38}
\,\left[W^{B}_{m,r},\,L_{\perp},\,H \right]&=[-m]_{\mathcal{R}(p,q)}\,H\,W^{B}_{m,r}\nonumber\\
\,\left[\bar{W}^{B}_{m,r},\,L_{\perp},\,H\right]&=[-m]_{\mathcal{R}(p,q)}\,H\,\bar{W}^{B}_{m,r}\nonumber\\
\,\left[W^{F}_{m,r},\,L_{\perp},\, H\right]&=[-\big(m-\frac{1}{2}\big)]_{\mathcal{R}(p,q)}\,H\,W^{F}_{m,r}\nonumber\\
\,\left[\bar{W}^{F}_{m,r},\,L_{\perp},\, H\right]&=[-\big(m+\frac{1}{2}\big)]_{\mathcal{R}(p,q)}\,H\,\bar{W}^{F}_{m,r}.
\end{align}

From the relation \eqref{eq38}, we note that these $\mathcal{R}(p,q)$ super 3-brackets are
not closed. But for the bosonic $\mathcal{R}(p,q)$-operators $W^{B}_{0,r}$ and $\bar{W}^{B}_{0,r}$,  we have null 3-commutators:
\begin{align}\label{eq39}
\left[W^{B}_{0,r},\,L_{\perp},\,H \right]=\left[\bar{W}^{B}_{0,r},\,L_{\perp},\,H \right]=0.
\end{align}
It should be pointed out that for the fermionic ones, there
do not exist such kinds of null 3-brackets.

Introducing  the $\mathcal{R}(p,q)$-operators \eqref{eq28} and \eqref{eq29} into the $\mathcal{R}(p,q)$ super $3$-bracket \eqref{eq5}, we can determine the $\mathcal{R}(p,q)$ super $W_{1+\infty}$ 3-algebra.
Due to the many of relations of the deformed 3-algebra, we focus our study on the
$\mathcal{R}(p,q)$-operators $W^{B}_{m,r}$ and $W^{F}_{m,r}$. Then, we obtain the following $\mathcal{R}(p,q)$ super-sub-3-algebra
\begin{align}\label{eq40}
\left[W^{B}_{m_1,r_1},\,W^{B}_{m_2,r_2},\,W^{B}_{m_3,r_3}\right]&=K(P,Q)\bigg[f^{m_2,r_2}_{m_3,r_3}p^{\alpha_2}\sum_{\alpha_2=0}^{r_1-1}q^{\bar{\alpha}}\mathbb{C}^{\alpha_2}_{r_1-1}\,\mathbb{A}^{\alpha_2}_{m_2+m_3+r_2+r_3-2-\alpha_1}  \nonumber\\&-f^{m_1,r_1}_{m_3,r_3}p^{\alpha_2}\sum_{\alpha_2=0}^{r_2-1}q^{\tilde{\alpha}}\mathbb{C}^{\alpha_2}_{r_2-1}\,\mathbb{A}^{\alpha_2}_{m_1+m_3+r_2+r_3-2-\alpha_1}\nonumber\\&+ f^{m_1,r_1}_{m_2,r_2}p^{\alpha_2}\sum_{\alpha_2=0}^{r_3-1}q^{\hat{\alpha}}\mathbb{C}^{\alpha_2}_{r_3-1}\,\mathbb{A}^{\alpha_2}_{m_1+m_3+r_1+r_2-2-\alpha_1} \bigg]W^{B}_{\tilde{m},\tilde{r}-2-\alpha_1-\alpha_2},
\end{align}
\begin{align}\label{eq41}
\left[W^{B}_{m_1,r_1},\,W^{B}_{m_2,r_2},\,W^{B}_{m_3,r_3}\right]&=K(P,Q)\bigg[f^{m_2,r_2}_{m_3,r_3}p^{\alpha_2}\sum_{\alpha_2=0}^{r_1-1}q^{\bar{\alpha}}\mathbb{C}^{\alpha_2}_{r_1-1}\,\mathbb{A}^{\alpha_2}_{m_2+m_3+r_2+r_3-2-\alpha_1}  \nonumber\\&-f^{m_1,r_1}_{m_3,r_3}p^{\alpha_2}\sum_{\alpha_2=0}^{r_2-1}q^{\tilde{\alpha}}\mathbb{C}^{\alpha_2}_{r_2-1}\,\mathbb{A}^{\alpha_2}_{m_1+m_3+r_2+r_3-2-\alpha_1}\nonumber\\&+ f^{m_1,r_1}_{m_2,r_2}p^{\alpha_2}\sum_{\alpha_2=0}^{r_3-1}q^{\hat{\alpha}}\mathbb{C}^{\alpha_2}_{r_3-1}\,\mathbb{A}^{\alpha_2}_{m_1+m_3+r_1+r_2-2-\alpha_1} \bigg]W^{F}_{\tilde{m},\tilde{r}-2-\alpha_1-\alpha_2},
\end{align}
\begin{eqnarray}\label{eq41a}
\left[W^{B}_{m_1,r_1},\,W^{F}_{m_2,r_2},\,W^{F}_{m_3,r_3}\right]=\left[W^{F}_{m_1,r_1},\,W^{F}_{m_2,r_2},\,W^{F}_{m_3,r_3}\right]=0,
\end{eqnarray}
where $\bar{\alpha}=(r_1-1-\alpha_2)(m_2+m_3+r_2+r_3-2-\alpha_1),\,\tilde{\alpha}=(r_2-1-\alpha_2)(m_1+m_3+r_2+r_3-2-\alpha_1),$ $\hat{\alpha}=(r_3-1-\alpha_2)(m_1+m_3+r_1+r_2-2-\alpha_1),$ $\tilde{m}=m_1+m_2+m_3$ and $\tilde{r}=r_1+r_2+r_3.$

From the relations obtained  in the previous section, the GSBI
\eqref{eq16}  holds for the $\mathcal{R}(p,q)$ super $W_{1+\infty}$ sub-3-algebra given by the relations  \eqref{eq40},\eqref{eq41},and \eqref{eq41a}. However
the relations \eqref{eq40} and \eqref{eq41} do not obey the GSJI \eqref{eq8} and super FI\cite{CK}:
\begin{align}\label{eq42}
\left[A,B,\left[C,D,E\right]\right]&=\left[\left[A,B,C\right],D,E\right]+(-1)^{|C|\big(|A|+|B|\big)}\left[C,\left[A,B,D\right],E\right]\nonumber\\ &+(-1)^{\big(|C|+|D|\big)\big(|A|+|B|\big)}\left[C,D\left[A,B,E\right]\right].
\end{align}

For the $\mathcal{R}(p,q)$-operators $W^{B(F)}_{m,2}$ and $W^{B(F)}_{m,1}$, we observe that
they yield the $\mathcal{R}(p,q)$-super Virasoro-Witt algebra:
\begin{align}\label{eq43}
\,\left[W^{B}_{m_1,2},W^{B(F)}_{m_2,2}\right]&=\big([m_2]_{\mathcal{R}(p,q)}-[m_1]_{\mathcal{R}(p,q)}\big)W^{B(F)}_{m_1+m_2,2},\nonumber\\
\,\left[W^{B}_{m_1,2},W^{B(F)}_{m_2,1}\right]&=[m_2]_{\mathcal{R}(p,q)}\,W^{B(F)}_{m_1+m_2,1},\nonumber\\
\,\left[W^{B}_{m_1,1},W^{F}_{m_2,2}\right]&=[-m_1]_{\mathcal{R}(p,q)}\,W^{F}_{m_1+m_2,1},
\end{align} 
other (anti) commutators are zeros. Moreover these operators
also generate the $\mathcal{R}(p,q)$-super Witt 3-algebra:
\begin{align}\label{eq44}
\,\left[W^{B}_{m_1,2},\,W^{B}_{m_2,2},\,W^{B(F)}_{m_3,1}\right]&=\big([m_2]_{\mathcal{R}(p,q)}-[m_1]_{\mathcal{R}(p,q)}\big)\big(W^{B(F)}_{m_1+m_2+m_3,2}\nonumber\\&+[m_3]_{\mathcal{R}(p,q)}\,W^{B(F)}_{m_1+m_2+m_3,1}\big),\nonumber\\
\,\left[W^{B}_{m_1,2},\,W^{B}_{m_2,1},\,W^{B(F)}_{m_3,1}\right]&=\big([m_2]_{\mathcal{R}(p,q)}-[m_3]_{\mathcal{R}(p,q)}\big)W^{B(F)}_{m_1+m_2+m_3,1},
\end{align} 
other 3-(anti)commutators are zeros. As a sub-3-algebra
of the relations \eqref{eq40},\eqref{eq41}, and \eqref{eq41a}, the forms of the  $\mathcal{R}(p,q)$-super Witt 3-algebra
\eqref{eq44} are simple. Although the GSBI \eqref{eq8} is guaranteed to hold for the relation \eqref{eq44}, we find that the GSJI \eqref{eq8} and super FI
\eqref{eq42} still fail for it.

Let us now investigate the case of the deformed  $n$-algebra. Putting the $\mathcal{R}(p,q)$-operators \eqref{eq28} and \eqref{eq29} into the $\mathcal{R}(p,q)$ super $n$-bracket \eqref{eq1},  we may derive the $\mathcal{R}(p,q)$ super $W_{1+\infty}$ $n$-algebra.
However the form of the higher order structure constants
appears to become more complicated due to the mixed
product of the bosonic and fermionic generators, it is hard
to give the all explicit expressions of multibrackets.

 For
this reason, here we only focus on the $\mathcal{R}(p,q)$-operators $W^B_{m,r}$
and $W^F_{m,r}$. Performing straightforward computations, we
obtain the $\mathcal{R}(p,q)$ super $n$-algebra
\begin{align}\label{eq45}
\left[W^{B}_{m_1,r_1},\cdots,W^{B}_{m_{n-1},r_{n-1}}, W^{B(F)}_{m_n,r_n}\right]&=\frac{K(P,Q)}{2^{\alpha}}\bigg(\frac{[-2\sum_{l=1}^{n}m_{l}]_{\mathcal{R}(p,q)}}{[-\sum_{l=1}^{n}m_{l}]_{\mathcal{R}(p,q)}}\bigg)^{\alpha}\epsilon_{1 2 \cdots n}^{i_1 i_2 \cdots i_n}\nonumber\\&\times \sum_{\alpha_1=0}^{\beta_1}
\sum_{\alpha_2=0}^{\beta_2}
\cdots \sum_{\alpha_{n-1}=0}^{\beta_{n-1}}
\mathbb{C}_{\beta_1}^{\alpha_1}\cdots \mathbb{C}_{\beta_{n-1}}^{\alpha_{n-1}}
\cdots \mathbb{A}_{m_{i_2}+r_{i_2}-1}^{\alpha_1}\nonumber\\
&\times \mathbb{A}_{m_{i_3}+r_{i_3}-1}^{\alpha_2}\cdots \mathbb{A}_{m_{i_n}+r_{i_n}-1}^{\alpha_{n-1}}\,q^{\bar{\lambda}}p^{\tilde{\lambda}}
W_{\bar{m},\bar{r}}^{B(F)},
\end{align}
other n-(anti)commutators are zeros, where $\bar{m}=\sum_{i=1
}^{n}m_i,\,\bar{r}=\sum_{i=1
}^{n}r_i-\sum_{i=1}^{n-1}\alpha_i-(n-1)$ and 
\begin{eqnarray}\label{eq46}
\beta_k=\left\{
\begin{array}{cc}
r_{i_1}-1,& k=1, \\
\sum_{j=1}^k r_{i_j}-k-\sum_{i=1}^{k-1}\alpha_i,   & 2\leqslant k\leqslant n-1.\\
\end{array}\right.
\end{eqnarray}

From  the associativity of the $\mathcal{R}(p,q)$-operators, the $\mathcal{R}(p,q)$-super $W_{1+\infty}$ $n$-algebra
\eqref{eq45} with $n$ even is the  generalized Lie algebra which satisfies the GSJI \eqref{eq8}. When n is odd, the
GSBI \eqref{eq16} is guaranteed to hold for it.
It is known that there exists a nontrivial sub-$2$-$n$-algebra
in the $W_{1+\infty}$ $2n$-algebra \cite{ZDYWZ}.

 Now, we investigate the
Supersymmetric case.  Setting the correspondance of the $\mathcal{R}(p,q)$-operators
$W_{m,n+1}^{B(F)}\rightarrow Q^{-\frac{1}{2n-1}}W_{m,n+1}^{B(F)}$,
with the coefficient
$Q$  given by:
\begin{eqnarray}\label{eq48}
Q=\sum_{(\alpha_1,\alpha_2,\cdots,\alpha_{2n-1})\in S_{2n-1} }\mathbb{C}_{\beta_1}^{\alpha_1}
\mathbb{C}_{\beta_2}^{\alpha_2}\cdots  \mathbb{C}_{\beta_{2n-1}}^{\alpha_{2n-1}}
\epsilon_{1 2 \cdots {(2n-1)}}^{\alpha_1 \alpha_2 \cdots \alpha_{2n-1}}, \end{eqnarray}
$S_{2n-1}$ is the permutation group of $\{1,2,\cdots,2n-1\}$. Thus, 
  the $\mathcal{R}(p,q)$-operators $W^{B}_{m,n+1}$ and $W^{F}_{m,n+1}$
 with any index $n + 1 \geq 2$, satisfy the $\mathcal{R}(p,q)$ super sub-2n-algebra:
\begin{eqnarray}\label{eq47}
\left[W^{B}_{m_1,n+1},\cdots,W^{B}_{m_{2n-1},n+1}, W^{B(F)}_{m_{2n},n+1}\right]=\frac{\mathbb{V}_{2n}}{2}\bigg(\frac{[-2\sum_{l=1}^{2n}m_{l}]_{\mathcal{R}(p,q)}}{[-\sum_{l=1}^{2n}m_{l}]_{\mathcal{R}(p,q)}}\bigg)W^{B(F)}_{\sum_{i=1}^{2n}m_i,n+1},
\end{eqnarray}
where 
   $\mathbb{V}_{2n}$
 is the Vandermonde determinant
 \begin{eqnarray}\label{eq49}
 \mathbb{V}_{2n}=\prod_{1\leq j<k\leq 2n}\big([m_k]_{\mathcal{R}(p,q)}-[m_j]_{\mathcal{R}(p,q)}\big)
 \end{eqnarray}
 and other $2n$-(anti)commutators are zeros.
 
 Note that the $\mathcal{R}(p,q)$ super $W_{1+\infty}$ sub-2n-algebra \eqref{eq47} satisfies the GSJI \eqref{eq8}, however it can be shown that the relation \eqref{eq47} does
 not satisfy the super FI\cite{CK}:
 \begin{small}
 \begin{align}\label{eq50}
 \left[A_1,\cdots,A_{n-1}\left[A_n,\cdots,A_{2n-1}\right]\right]&=\sum_{i=0}^{n-1}(-1)^{\big(|A_n|+\cdots+|A_{n-1+i}|\big)\big(|A_1|+\cdots+|A_{n-1}|\big)}\nonumber\\&\times \left[A_n,\cdots,A_{n-1+i}\left[A_1,\cdots,A_{n-1}, A_{n+i}\right]A_{n+i+1},\cdots,A_{2n-1}\right].
 \end{align}
  \end{small}
 Thus the relation \eqref{eq47} is a generalized  quantum super Lie algebra.
 \begin{example}
 	 For examples, we derive the first two $\mathcal{R}(p,q)$
 super $W_{1+\infty}$ sub-2n-algebras.
 \begin{enumerate}
 	\item[(i)] For the $\mathcal{R}(p,q)$-operators $W^{B}_{m,3}$ and $W^{F}_{m,3}$,  we have the $\mathcal{R}(p,q)$-super sub-4-algebra
 	\begin{align*}
 	\left[W^{B}_{m_1,3},W^{B}_{m_2,3},W^{B}_{m_3,3},W^{B(F)}_{m_4,3}\right]&=\mathcal{A}(p,q)\big([m_4]_{\mathcal{R}(p,q)}-[m_3]_{\mathcal{R}(p,q)}\big)\big([m_4]_{\mathcal{R}(p,q)}-[m_2]_{\mathcal{R}(p,q)}\big)\nonumber\\&\times \big([m_4]_{\mathcal{R}(p,q)}-[m_1]_{\mathcal{R}(p,q)}\big)\big([m_3]_{\mathcal{R}(p,q)}-[m_2]_{\mathcal{R}(p,q)})\nonumber\\&\times\big([m_3]_{\mathcal{R}(p,q)}-[m_1]_{\mathcal{R}(p,q)}\big)\big([m_2]_{\mathcal{R}(p,q)}-[m_1]_{\mathcal{R}(p,q)}\big)W^{B(F)}_{\sum_{i=1}^{4}m_i,3},
 	\end{align*}
 	where $\mathcal{A}(p,q)=\bigg(\frac{[-2\sum_{l=1}^{4}m_{l}]_{\mathcal{R}(p,q)}}{[-\sum_{l=1}^{4}m_{l}]_{\mathcal{R}(p,q)}}\bigg)$ and other $4$-(anti)commutators are zero.
 	\item[(ii)] For the $\mathcal{R}(p,q)$-operators $W^{B}_{m,4}$ and $W^{F}_{m,4}$,  we have the $\mathcal{R}(p,q)$-super sub-$6$-algebra
 	\begin{align*}
 	\left[W^{B}_{m_1,4},\cdots,W^{B}_{m_5,4},W^{B(F)}_{m_6,4}\right]&=\bar{\mathcal{A}}(p,q)\prod_{1\leq j<k\leq 6}\bigg([m_k]_{\mathcal{R}(p,q)}-[m_j]_{\mathcal{R}(p,q)}\bigg)W^{B(F)}_{\sum_{i=1}^{6}m_i,4},
 	\end{align*}
 	where $\bar{\mathcal{A}}(p,q)=\bigg(\frac{[-2\sum_{l=1}^{6}m_{l}]_{\mathcal{R}(p,q)}}{[-\sum_{l=1}^{6}m_{l}]_{\mathcal{R}(p,q)}}\bigg)$ and other 6-(anti)commutators are zero.
 \end{enumerate}
\end{example}
\section{super $W_{1+\infty}$ $n$-algebra,  Landau Problem and quantum algebras}
We deduce particular cases of super $n$-algebra and super $W_{1+\infty}$ $n$-algebra with the Landau problem induced by some quantum algebras.
\begin{enumerate}
	\item[(i)] Taking $\mathcal{R}(x)=\frac{x-1}{q-1},$ we obtain : the $q$-super
	multibracket of order $n$ is given by:
	\begin{align*}
	\left[A_1,A_2,\cdots,A_n\right]=\frac{1}{2^{\alpha}}\bigg(\frac{[-2\sum_{l=1}^{n}m_{l}]_{q}}{[-\sum_{l=1}^{n}m_{l}]_{q}}\bigg)^{\alpha}\epsilon_{1 2 \cdots n}^{i_1 i_2 \cdots i_n}\big(-1\big)^{\sum_{k=1}^{n-1}|A_k|(\sum_{l=k+1,i_l< i_k}^{n}|A_{i_l}|)}A_{i_1}\,A_{i_2}\cdots A_{i_n},
	\end{align*}
	with
	\begin{eqnarray*}
	[n]_{q}=\frac{1-q^n}{1-q}, \quad n\in\mathbb{N}.
	\end{eqnarray*}    
	For the bosonic operators, we have the 
	$n$-commutator:
	\begin{eqnarray*}
	\left[A_1,A_2,\cdots,A_n\right]=\frac{1}{2^{\alpha}}\bigg(\frac{[-2\sum_{l=1}^{n}m_{l}]_{q}}{[-\sum_{l=1}^{n}m_{l}]_{q}}\bigg)^{\alpha}\epsilon_{1 2 \cdots n}^{i_1 i_2 \cdots i_n}\,A_{i_1}\,A_{i_2}\cdots A_{i_n}.
	\end{eqnarray*}
	 For $n=2,$ we deduce the $q$-Lie super bracket.
	\begin{eqnarray*}
	\left[A_1,A_2\right]=\frac{1}{2}\bigg(\frac{[-2\big(m_{1}+m_{2}\big)]_{q}}{[-\big(m_{1}+m_{2}\big)]_{q}}\bigg)\bigg(A_1\,A_2-(-1)^{|A_2||A_1|}A_2\,A_1\bigg).
	\end{eqnarray*}
	From the $q$-super multibracket, we obtain the $q$-skewsymmetry
	\begin{align}
	\left[A_{\sigma_1},A_{\sigma_2},\cdots,A_{\sigma_n}\right]&=\frac{1}{2^{\alpha}}\bigg(\frac{[-2\sum_{l=1}^{n}m_{l}]_{q}}{[-\sum_{l=1}^{n}m_{l}]_{q}}\bigg)^{\alpha}\epsilon_{1 2 \cdots n}^{i_1 i_2 \cdots i_n}\nonumber\\&\times\big(-1\big)^{\sum_{k=1}^{n-1}|A_{\sigma_k}|(\sum_{l=k+1,i_l< i_k}^{n}|A_{\sigma_l}|)}\left[A_{1},A_{2},\cdots,A_{n}\right]
	\end{align}
	and the following commutators:
	\begin{align*}
	\left[B,A_{1},A_{2},\cdots A_{n-1}\right]&=\frac{1}{2^{\alpha}}\bigg(\frac{[-2\sum_{l=1}^{n}m_{l}]_{q}}{[-\sum_{l=1}^{n}m_{l}]_{q}}\bigg)^{\alpha}\sum_{j=0}^{n-1}(-1)^{j}\epsilon_{12  \cdots (n-1)}^{i_1i_2 \cdots i_{n-1}}\nonumber\\&\times(-1)^{|p_1|}A_{i_1},A_{i_2},\cdots,A_{i_{j}}BA_{i_{j+1}}\cdots A_{i_{n-1}},
	\end{align*}
	\begin{align*}
	\left[B,A_{1},A_{2},\cdots A_{n},Z\right]&=\frac{1}{2^{\alpha}}\bigg(\frac{[-2\sum_{l=1}^{n}m_{l}]_{q}}{[-\sum_{l=1}^{n}m_{l}]_{q}}\bigg)^{\alpha}\bigg(\sum_{j=0}^{n}(-1)^{j}\sum_{k=0}^{n-j}(-1)^k\epsilon_{12  \cdots n}^{i_1i_2 \cdots i_{n}}\nonumber\\&\times(-1)^{|p_2|} A_{i_1},\cdots,A_{i_{k}}BA_{i_{k+1}}\cdots A_{i_{n-j}}ZA_{i_{n-j+1}}\cdots A_{i_{n}}\nonumber\\
	&-\sum_{j=0}^{n}(-1)^{j}\sum_{k=0}^{n-j}(-1)^k\epsilon_{12  \cdots n}^{i_1i_2 \cdots i_{n}}\nonumber\\&\times(-1)^{|p_3|} A_{i_1},\cdots,A_{i_{k}}ZA_{i_{k+1}}\cdots A_{i_{n-j}}BA_{i_{n-j+1}}\cdots A_{i_{n}}\bigg),
	\end{align*}
	for any fixed $B$, $Z$, and $p_i,\,i\in\{1,2,3\}$ are given by the relations \eqref{eq13}, \eqref{eq14} and \eqref{eq15}.
	
	The $q$-deformed of the operators \eqref{eq28} and \eqref{eq29} satisfy 
	the $q$-super $W_{1+\infty}$ algebra governed by the commutation relations \eqref{eq30},\eqref{eq31}, \eqref{eq32} and 
	\begin{align*}
	\,\left[\bar{W}^{B}_{m_1,r_1},\,W^{F}_{m_2,r_2}\right]&=-\sum_{\alpha_1=0}^{r_2-1}q^{(r_2-1-\alpha_1)(m_1+r_1-1)}\mathbb{C}^{\alpha_1}_{r_2-1}\,\mathbb{A}^{\alpha_1}_{m_1+r_1-1}\,W^{F}_{\bar{m},\bar{r}},\\
	\,\left[\bar{W}^{B}_{m_1,r_1},\,\bar{W}^{F}_{m_2,r_2}\right]&=\sum_{\alpha_1=0}^{r_1-1}q^{(r_1-1-\alpha_1)(m_2+r_2-1)}\,\mathbb{C}^{\alpha_1}_{r_1-1}\,\mathbb{A}^{\alpha_1}_{m_2+r_2-1}\,\bar{W}^{F}_{\bar{m},\bar{r}},\\
	\,\left[W^{F}_{m_1,r_1},\bar{W}^{F}_{m_2,r_2}\right]&=\sum_{\alpha_1=0}^{r_1-1}q^{(r_1-1-\alpha_1)(m_2+r_2-1)}\mathbb{C}^{\alpha_1}_{r_1-1}\mathbb{A}^{\alpha_1}_{m_2+r_2-1}W^{B}_{\bar{m},\bar{r}}-f^{m_1,r_1}_{m_2,r_2}\bar{W}^{B}_{\bar{m},\bar{r}},
	\end{align*}
	where $\bar{m}=m_1+m_2,$ $\bar{r}=r_1+r_2-1-\alpha_1,$
	and \begin{eqnarray}
	\mathbb{C}^{k}_{n} := \frac{[n]!_{q}}{[k]!_{q}[n-k]!_{q}},\quad  n\geq k,\quad 	\mathbb{A}_n^k:=
	\left\{\begin{array}{cc}
	[n]_{q}[n-1]_{q}
	\cdots[n-k+1]_{q},& k\leqslant n,\\
	0,                  &k>n.\end{array}\right. 
	\end{eqnarray}
\begin{align*}
	f^{m_1,r_1}_{m_2,r_2}&=\bigg(\sum_{\alpha_1=0}^{r_1-1}q^{(r_1-1-\alpha_1)(m_2+r_2-1)}\mathbb{C}^{\alpha_1}_{r_1-1}\mathbb{A}^{\alpha_1}_{m_2+r_2-1}\nonumber\\&-\sum_{\alpha_1=0}^{r_1-1}q^{(r_2-1-\alpha_1)(m_1+r_1-1)}\mathbb{C}^{\alpha_1}_{r_2-1}\mathbb{A}^{\alpha_1}_{m_1+r_1-1}\bigg).
\end{align*}
	Besides, the commutation
	relations between the $q$-operators \eqref{eq28} and \eqref{eq29} with the angular momentum are given as follows:
	\begin{align*}
	\,\left[L_{\perp},\,W^{B}_{m,r}\right]&=[m]_{q}\,W^{B}_{m,r}\nonumber\\
	\,\left[L_{\perp},\,\bar{W}^{B}_{m,r}\right]&=[m]_{q}\,\bar{W}^{B}_{m,r}\nonumber\\
	\,\left[L_{\perp},\,W^{F}_{m,r}\right]&=[m-\frac{1}{2}]_{q}\,W^{F}_{m,r}\nonumber\\
	\,\left[L_{\perp},\,\bar{W}^{F}_{m,r}\right]&=[m+\frac{1}{2}]_{q}\,\bar{W}^{F}_{m,r}.
	\end{align*}
	 Thus, we deduce the $q$-super 3-brackets by:
	\begin{align*}
	\,\left[W^{B}_{m,r},\,L_{\perp},\,H \right]&=[-m]_{q}\,H\,W^{B}_{m,r}\nonumber\\
	\,\left[\bar{W}^{B}_{m,r},\,L_{\perp},\,H\right]&=[-m]_{q}\,H\,\bar{W}^{B}_{m,r}\nonumber\\
	\,\left[W^{F}_{m,r},\,L_{\perp},\, H\right]&=[-\big(m-\frac{1}{2}\big)]_{q}\,H\,W^{F}_{m,r}\nonumber\\
	\,\left[\bar{W}^{F}_{m,r},\,L_{\perp},\, H\right]&=[-\big(m+\frac{1}{2}\big)]_{q}\,H\,\bar{W}^{F}_{m,r}.
	\end{align*}
	Moreover, the $q$-super-sub-3-algebra is generated by:
	\begin{align*}
	\left[W^{B}_{m_1,r_1},\,W^{B}_{m_2,r_2},\,W^{B}_{m_3,r_3}\right]&=\bigg[f^{m_2,r_2}_{m_3,r_3}\sum_{\alpha_2=0}^{r_1-1}q^{\bar{\alpha}}\mathbb{C}^{\alpha_2}_{r_1-1}\,\mathbb{A}^{\alpha_2}_{m_2+m_3+r_2+r_3-2-\alpha_1}  \nonumber\\&-f^{m_1,r_1}_{m_3,r_3}\sum_{\alpha_2=0}^{r_2-1}q^{\tilde{\alpha}}\mathbb{C}^{\alpha_2}_{r_2-1}\,\mathbb{A}^{\alpha_2}_{m_1+m_3+r_2+r_3-2-\alpha_1}\nonumber\\&+ f^{m_1,r_1}_{m_2,r_2}\sum_{\alpha_2=0}^{r_3-1}q^{\hat{\alpha}}\mathbb{C}^{\alpha_2}_{r_3-1}\,\mathbb{A}^{\alpha_2}_{m_1+m_3+r_1+r_2-2-\alpha_1} \bigg]W^{B}_{\tilde{m},\tilde{r}-2-\alpha_1-\alpha_2}	\end{align*}
	\begin{align*}
	\left[W^{B}_{m_1,r_1},\,W^{B}_{m_2,r_2},\,W^{B}_{m_3,r_3}\right]&=\bigg[f^{m_2,r_2}_{m_3,r_3}\sum_{\alpha_2=0}^{r_1-1}q^{\bar{\alpha}}\mathbb{C}^{\alpha_2}_{r_1-1}\,\mathbb{A}^{\alpha_2}_{m_2+m_3+r_2+r_3-2-\alpha_1}  \nonumber\\&-f^{m_1,r_1}_{m_3,r_3}\sum_{\alpha_2=0}^{r_2-1}q^{\tilde{\alpha}}\mathbb{C}^{\alpha_2}_{r_2-1}\,\mathbb{A}^{\alpha_2}_{m_1+m_3+r_2+r_3-2-\alpha_1}\nonumber\\&+ f^{m_1,r_1}_{m_2,r_2}\sum_{\alpha_2=0}^{r_3-1}q^{\hat{\alpha}}\mathbb{C}^{\alpha_2}_{r_3-1}\,\mathbb{A}^{\alpha_2}_{m_1+m_3+r_1+r_2-2-\alpha_1} \bigg]W^{F}_{\tilde{m},\tilde{r}-2-\alpha_1-\alpha_2},
	\end{align*}
	\begin{eqnarray*}
	\left[W^{B}_{m_1,r_1},\,W^{F}_{m_2,r_2},\,W^{F}_{m_3,r_3}\right]=\left[W^{F}_{m_1,r_1},\,W^{F}_{m_2,r_2},\,W^{F}_{m_3,r_3}\right]=0,
	\end{eqnarray*}
	For the $q$-operators $W^{B(F)}_{m,2}$ and $W^{B(F)}_{m,1}$, we observe that
	they yield the $q$-super Witt algebra:
	\begin{align*}
	\,\left[W^{B}_{m_1,2},W^{B(F)}_{m_2,2}\right]&=\big([m_2]_{q}-[m_1]_{q}\big)W^{B(F)}_{m_1+m_2,2},\nonumber\\
	\,\left[W^{B}_{m_1,2},W^{B(F)}_{m_2,1}\right]&=[m_2]_{q}\,W^{B(F)}_{m_1+m_2,1},\nonumber\\
	\,\left[W^{B}_{m_1,1},W^{F}_{m_2,2}\right]&=[-m_1]_{q}\,W^{F}_{m_1+m_2,1},
	\end{align*} 
	other (anti) commutators are zeros. Moreover, we get the $q$-super Witt 3-algebra:
	\begin{align*}
	\,\left[W^{B}_{m_1,2},\,W^{B}_{m_2,2},\,W^{B(F)}_{m_3,1}\right]&=\big([m_2]_{q}-[m_1]_{q}\big)\big(W^{B(F)}_{m_1+m_2+m_3,2}\nonumber\\&+[m_3]_{q}\,W^{B(F)}_{m_1+m_2+m_3,1}\big),\nonumber\\
	\,\left[W^{B}_{m_1,2},\,W^{B}_{m_2,1},\,W^{B(F)}_{m_3,1}\right]&=\big([m_2]_{q}-[m_3]_{q}\big)W^{B(F)}_{m_1+m_2+m_3,1},
	\end{align*} 
	other 3-(anti)commutators are zeros.
	The $q$-super $n$-algebra is given by the $n$-commutator:
	\begin{align*}
	\left[W^{B}_{m_1,r_1},\cdots,W^{B}_{m_{n-1},r_{n-1}}, W^{B(F)}_{m_n,r_n}\right]&=\frac{1}{2^{\alpha}}\bigg(\frac{[-2\sum_{l=1}^{n}m_{l}]_{q}}{[-\sum_{l=1}^{n}m_{l}]_{q}}\bigg)^{\alpha}\epsilon_{1 2 \cdots n}^{i_1 i_2 \cdots i_n}\nonumber\\&\times \sum_{\alpha_1=0}^{\beta_1}
	\sum_{\alpha_2=0}^{\beta_2}
	\cdots \sum_{\alpha_{n-1}=0}^{\beta_{n-1}}
	\mathbb{C}_{\beta_1}^{\alpha_1}\cdots \mathbb{C}_{\beta_{n-1}}^{\alpha_{n-1}}
	\cdots \mathbb{A}_{m_{i_2}+r_{i_2}-1}^{\alpha_1}\nonumber\\
	&\times \mathbb{A}_{m_{i_3}+r_{i_3}-1}^{\alpha_2}\cdots \mathbb{A}_{m_{i_n}+r_{i_n}-1}^{\alpha_{n-1}}\,q^{\bar{\lambda}}
	W_{\bar{m},\bar{r}}^{B(F)},
	\end{align*}
	other n-(anti)commutators are zeros.
	Also, we have the $q$ super sub-2n-algebra:
	\begin{align*}
	\left[W^{B}_{m_1,n+1},\cdots,W^{B}_{m_{2n-1},n+1}, W^{B(F)}_{m_{2n},n+1}\right]&=\frac{1}{2}\bigg(\frac{[-2\sum_{l=1}^{2n}m_{l}]_{q}}{[-\sum_{l=1}^{2n}m_{l}]_{q}}\bigg)\nonumber\\&\times\prod_{1\leq j<k\leq 2n}\big([m_k]_{q}-[m_j]_{q}\big)W^{B(F)}_{\sum_{i=1}^{2n}m_i,n+1},
	\end{align*}
	and other $2n$-(anti)commutators are zeros. For the $q$-operators $W^{B}_{m,3}$ and $W^{F}_{m,3}$,  we have the $q$-super sub-4-algebra
	\begin{align*}
	\left[W^{B}_{m_1,3},W^{B}_{m_2,3},W^{B}_{m_3,3},W^{B(F)}_{m_4,3}\right]&=\frac{1}{2}\bigg(\frac{[-2\sum_{l=1}^{4}m_{l}]_{q}}{[-\sum_{l=1}^{4}m_{l}]_{q}}\bigg)\big([m_4]_{q}-[m_3]_{q}\big)\big([m_4]_{q}-[m_2]_{q}\big)\nonumber\\&\times \big([m_4]_{q}-[m_1]_{q}\big)\big([m_3]_{q}-[m_2]_{q})\nonumber\\&\times\big([m_3]_{q}-[m_1]_{q}\big)\big([m_2]_{q}-[m_1]_{q}\big)W^{B(F)}_{\sum_{i=1}^{4}m_i,3},
	\end{align*}
	 and other 4-(anti)commutators are zeros.
	Moreover,  the $q$-super sub-6-algebra is given by: 
	\begin{eqnarray*}
	\left[W^{B}_{m_1,4},\cdots,W^{B}_{m_5,4},W^{B(F)}_{m_6,4}\right]=\frac{1}{2}\bigg(\frac{[-2\sum_{l=1}^{6}m_{l}]_{q}}{[-\sum_{l=1}^{6}m_{l}]_{q}}\bigg)\prod_{1\leq j<k\leq 6}\big([m_k]_{q}-[m_j]_{q}\big)W^{B(F)}_{\sum_{i=1}^{6}m_i,4},
	\end{eqnarray*}
	 and other 6-(anti)commutators are zeros.
	 \item[(i)] Setting $\mathcal{R}(x,y)=\frac{x-y}{p-q},$ we derive the $(p,q)$-super $W_{1+\infty}$ $n$-algebra and the Landau problem.
\end{enumerate}
\section{Concluding Remarks}
In this paper, the $\mathcal{R}(p,q)$-super $n$-bracket and properties like generalize super Jacobi identity (GSJI) have been analyzed and investigated. 
The $\mathcal{R}(p,q)$-super $W_{1+\infty}$ algebra and  the $\mathcal{R}(p,q)$-super $W_{1+\infty}$ $n$-algebra, which satisfies the (GSJI)  for  $n$
even have been presented and determined. Furthermore, the $\mathcal{R}(p,q)$-super $W_{1+\infty}$  sub-$2n$-algebra has been given and particular cases from known quantum algebras have been deduced. 
 More applications of this infinite-dimensional $n$-algebra in
physics should be of interest.
\section*{Acknowledgments}
FM was supported\footnote{``Funded by the Deutsche
	Forschungsgemeinschaft (DFG, German Research Foundation) -- Project ID
	541735537''} by an AIMS-DFG cooperation visit and by a Georg Forster
Fellowship of the Alexander von Humboldt Foundation.
RW was supported\footnote{``Funded by the Deutsche
	Forschungsgemeinschaft (DFG, German Research Foundation) --
	Project ID 427320536 -- SFB 1442, as well as under Germany's
	Excellence Strategy EXC 2044 390685587, Mathematics M\"unster:
	Dynamics -- Geometry -- Structure.''} by the Cluster of Excellence
\emph{Mathematics M\"unster} and the CRC 1442 \emph{Geometry:
	Deformations and Rigidity}.


\begin{thebibliography}{BDEG21}
	\bibitem{N}
	Y. Nambu, Generalized Hamiltonian dynamics, {\it Phys. Rev. D.} {\bf 7}, 2405 (1973).
	\bibitem{CFZ}
	T.L. Curtright, D.B. Fairlie and C.K. Zachos, {\it Ternary Virasoro-Witt algebra, Phys. Lett. B.}
	{\bf 666}, 386 (2008).
	\bibitem{CKJ}
	S. Chakrabortty, A. Kumar and S. Jain, $w_{\infty}$ 3-algebra, {\it  J. High Energy Phys.} {\bf 09}, 091 (2008) .
	\bibitem{CWZ}M. R. Chen, K. Wu and W. Z. Zhao, Super $w_{\infty}$ 3-algebra, {\it J. High Energy Phys.} {\bf 09}, 090, (2011).
	\bibitem{YDLZZ}S. K. Yao, L. Ding, P. Liu, C. H. Zhang, and W. Z. Zhao, On $d$-Dimensional Lattice (co)sine $n$-Algebra,
	{\it Commun. Theor. Phys.} {\bf 66}, 423 (2016).
	\bibitem{CWWWZ}M. R. Chen, S. K. Wang, X. L. Wang, K. Wu, and W. Z.
	Zhao, On $W_{1+\infty}$ 3-algebra and integrable system {\it Nucl. Phys. B.} {\bf 891}, 655 (2015).
	\bibitem{CTZ1}A. Cappelli, C. A. Trugenberger, and G. R. Zemba, Infinite symmetry in the quantum Hall effect,
	{\it Nucl. Phys. B.}
	{\bf 396}, 465-490, (1993).
	\bibitem{K} I. I. Kogan, Area-preserving diffeomorphisms, $W_{\infty}$ and ${\mathcal U}_q [{\rm sl}(2)]$ in Chern-Simons theory and the quantum Hall system, {\it Int. J. Mod. Phys. A}, {\bf 09}, 3887-3911 (1994).
	\bibitem{CTZ} A. Capelli, C. Trugenberger, and G. Zemba, Classification of quantum Hall universality classes by $W_1$+symmetry,  {\it Phys. Rev.
	Lett.} {\bf 72}, 1902 (1994).
	\bibitem{IKS}S. Iso, D. Karabali, and B. Sakita, Fermions in the lowest Landau level. Bosonization, W? algebra, droplets, chiral bosons,  {\it Phys. Lett. B.} {\bf 296}, 143-150
	(1992).
	\bibitem{A} H. Azuma, $W(\infty)$ algebra in the integer quantum Hall effects, {\it Prog. Theor. Phys.} {\bf 92}, 293-308 (1994).
	\bibitem{H1}K. Hasebe, Quantum Hall liquid on a noncommutative superplane,  {\it Phys. Rev. D}. {\bf 72}, 105017 (2005).
	\bibitem{ZDYWZ}C. H. Zhang, L. Ding, Z. W. Yan, K.Wu, and W. Z. Zhao, Super $W_{1+\infty}$-$n$-algebra in the Supersymmetric Landau Problem, {\it Commun. Theor. Phys.} {\bf 67}  648-654 (2017).
	\bibitem{HB1}M. N. Hounkonnou  and  J. D. Bukweli Kyemba,  $\mathcal{R}(p, q)$-deformed
	quantum algebras: coherent states and special functions, {\it J. Math. Phys.} {\bf 51}, 063518 
	(2010).
	\bibitem{CZ}T. Curtright and  C. Zachos, Deforming maps for quantum algebras, {\it Phys. Lett. B.} {\bf 243}, 237-244 (1990).
	\bibitem{WYLWZ}R. Wang, M. Li.  Yao, K.  Wu   and W. Zhao, On deformations of the Witt $n$-algebra, {\it J. Maths. Phys.} {\bf 59}, 103504  1-10  (2018).
	\bibitem{CJ}R.  Chakrabarti  and  R. Jagannathan,  A $(p,q)$-deformed Virasoro algebra, {\it J. Phys. A Math. Gen.} {\bf 25}, 2607-2614, (1992).
	\bibitem{HM}M. N. Hounkonnou  and F.  Melong,  $\mathcal{R}(p,q)$-deformed conformal Virasoro algebra, {\it J. Maths. Phys.}{\bf 60}, (2019).
	\bibitem{HMM} M. N. Hounkonnou, F.  Melong and M. Mitrovi\'c,  Generalized Witt, Witt $n$-algebras, Virasoro algebras and KdV
	equations induced from $\mathcal{R}(p,q)$-deformed quantum algebras, {\it Rev.  Math. Phys.} 2150011 (2020).
\bibitem{melongRaimar} F. Melong and R. Wulkenhaar, Characterization of quantum $W_{1+\infty}$-n-algebra and applications, (2024).
	\bibitem{melong2024} F. Melong, Conformal super Virasoro algebra: matrix model and quantum deformed algebra, {\it Rev. Math. Phys.} {\bf 36}, 2450006 (2023).
	\bibitem{melong2022}F. Melong, $\mathcal{R}(p,q)$-deformed super Virasoro $n$-algebras, {\it Rev. Math. Phys.} {\bf 33}, 2250038 (2022).
	\bibitem{melongwulkenhaar} F. Melong and R. Wulkenhaar, Generalized Heisenberg-Virasoro algebra and matrix models from quantum algebra, {\it J. Math. Phys.} {\bf 64}, 073505 (2023).
	\bibitem{HB}M. N. Hounkonnou  and J. D.  Kyemba Bukweli,  $\mathcal{R}(p,q)-$ calculus: differentiation and integration, {\it SUT Journal of Mathematics}, Vol {\bf 49}  (2), 145-167, (2013).
	\bibitem{TN}T. Nishino,    \emph{Function theory in several complex variables}, Translations of mathematical monographs, Volume  193 American Mathematical Society, Providence, Rhode Island (2001).
	\bibitem{HW}P. Hanlon and M. Wachs, On Lie k-algebras,  {\it Adv. Math.} {\bf 113}, 206 (1995).
	\bibitem{Br1}M. R. Bremner, Identities for the Ternary Commutator,  {\it J. Algebra} {\bf 206} 615, (1998).
	\bibitem{Br2} M. R. Bremner and L. A. Peresi, Ternary analogues of Lie and Malcev algebras, {\it Linear Algebra Appl.} {\bf 414} 1, (2006).
	\bibitem{CFJMZ}
	T. Curtright, D. Fairlie, X. Jin, L. Mezincescu and C. Zachos, Classical and quantal ternary algebras,
	{\it Phys. Lett. B.} {\bf 675}, 387 (2009) .
	\bibitem{DF}C. Devchand, D. Fairlie, J. Nuyts, and G. Weingart, Ternutator identities, {\it J.
	Phys. A: Math. Theor.} {\bf 42}, 475209 (2009).
	\bibitem{CK}
	N. Cantarini and V.G. Kac, Classification of simple linearly compact n-Lie superalgebras, arXiv:0909.3284.
	\bibitem{zz}
	C. Z. Zha and W. Z. Zhao, The $q$-deformation of super high-order Virasoro algebra,
	J. Math. Phys. {\bf 36} (1995) 967.
\end{thebibliography}
\end{document}